\documentclass[aps,prb,preprint,superscriptaddress]{revtex4-2}

\usepackage{graphicx}
\usepackage{dcolumn}
\usepackage{bm}
\usepackage[utf8]{inputenc}
\usepackage[T1]{fontenc}
\usepackage{mathptmx}
\usepackage{etoolbox}
\usepackage{multirow}
\usepackage{blindtext}
\usepackage{array}
\usepackage{amsmath}
\usepackage{microtype}
\usepackage{booktabs}
\usepackage{mhchem}
\usepackage{color}
\usepackage{url}

\begin{document}

\title{Isolating the Nonlinear Optical Response of a \ce{MoS2} Monolayer under Extreme Screening of a Metal Substrate}

\author{Tao Yang}
\email{tao.yang@uni-due.de}
 \affiliation{Faculty of Physics, University of Duisburg-Essen, 47057 Duisburg, Germany}
\author{Stephan Sleziona}
 \affiliation{Faculty of Physics, University of Duisburg-Essen, 47057 Duisburg, Germany}
\author{Erik Pollmann}
 \affiliation{Faculty of Physics, University of Duisburg-Essen, 47057 Duisburg, Germany}
\author{Eckart Hasselbrink}
 \affiliation{Faculty of Chemistry, University of Duisburg-Essen, 45117 Essen, Germany}
\author{Peter Kratzer}
 \affiliation{Faculty of Physics, University of Duisburg-Essen, 47057 Duisburg, Germany}
\author{Marika Schleberger}
 \affiliation{Faculty of Physics, University of Duisburg-Essen, 47057 Duisburg, Germany}
\author{R. Kramer Campen}
 \affiliation{Faculty of Physics, University of Duisburg-Essen, 47057 Duisburg, Germany}
\author{Yujin Tong}
\email{yujin.tong@uni-due.de}
 \affiliation{Faculty of Physics, University of Duisburg-Essen, 47057 Duisburg, Germany}

\date{\today}

\begin{abstract}
Transition metal dichalcogenides (TMDCs) monolayers, as two-dimensional (2D) direct bandgap semiconductors, hold promise for advanced optoelectronic and photocatalytic devices. Interaction with three-dimensional (3D) metals, like Au, profoundly affects their optical properties, posing challenges in characterizing the monolayer's optical responses within the semiconductor-metal junction. In this study, using precise polarization-controlled final-state sum frequency generation (FS-SFG), we successfully isolated the optical responses of a \ce{MoS2} monolayer from a \ce{MoS2}/Au junction. The resulting SFG spectra exhibit a linear lineshape, devoid of A or B exciton features, attributed to the strong dielectric screening and substrate induced doping. The linear lineshape illustrates the expected constant density of states (DOS) at the band edge of the 2D semiconductor, a feature often obscured by excitonic interactions in week-screening conditions such as in a free-standing monolayer. Extrapolation yields the onset of a direct quasiparticle bandgap of about $1.65\pm0.20$ eV, indicating a strong bandgap renormalization. This study not only enriches our understanding of the optical responses of a 2D semiconductor in extreme screening conditions but also provides a critical reference for advancing 2D semiconductor-based photocatalytic applications.

\end{abstract}


\maketitle

\section{Introduction}

The unique appeal of monolayers of transition metal dichalcogenides (TMDCs) lies in their distinctive characteristics as two-dimensional (2D) direct bandgap semiconductors \cite{mak_atomically_2010, splendiani_emerging_2010}. 
The strong light-matter interaction and the quantum confinement effect position these materials as promising candidates for advancing next-generation optoelectronic and photocatalytic devices \cite{wang_electronics_2012,mak_photonics_2016,mueller_exciton_2018,liu_promises_2021}. 
In light of these prospects, a fundamental question arises: How does a 2D TMDC monolayer maintain its semiconductor characteristics when interfaced with a three-dimensional (3D) metal? 
This query is pivotal, considering the well-known strong screening and doping effects associated with bulk metals. 
Beyond its fundamental implications, this question also holds practical relevance within the fields of TMDC-based 2D transistors \cite{radisavljevic_single-layer_2011, yin_single-layer_2012, bartolomeo_hysteresis_2017} and photocatalysis \cite{shi_temperature-mediated_2016,guo_enhanced_2020,strange_investigating_2020}. 
However, the precise characterization of the electronic properties of a 2D semiconductor on a 3D metal substrate presents substantial challenges. 
Techniques such as scanning tunneling spectroscopy (STS) offer a means of gauging the electronic bandgap by monitoring the onset of tunneling current \cite{bruix_single-layer_2016,silva_spatial_2022}. 
However, accurately determining the onset of tunneling current remains a non-trivial task, potentially introducing ambiguity in the derived bandgap energy \cite{wang_colloquium_2018}. 
An alternative approach involving time- and angle-resolved photoemission spectroscopy (tr-ARPES) shows promise for electronic bandgap determination \cite{cabo_observation_2015,liu_direct_2019}. 
However, the applicability of tr-ARPES is constrained by its demanding experimental requisites.

There is a well-established optical method in the device community, the \textit{Tauc plot}, which has been used extensively to straightforwardly determine the bandgap of 3D amorphous semiconductors \cite{tauc_optical_1968, klein_limitations_2023, grundmann_physics_2016}. 
While this method may also be adapted for use with crystalline semiconductors such as thin layers of TMDCs after certain modifications, challenges arise due to the presence of excitonic features at the band edge. 
There is an ongoing debate regarding the construction of the \textit{Tauc plot} for TMDC systems. Some studies involve drawing the slope from the lower-energy side of the A exciton \cite{zheng_photoconductivity_2020,paul_optical_2022}, while others base it on the quasiparticle bandgap \cite{mattinen_atomic_2017,wang_growth_2021} and simply ignore the excitonic features. Clearly, the accurate characterization of the onset of the band edge of a TMDC monolayer depends on a comprehensive understanding of its optical properties under different environments. 

Theoretically, a 2D semiconductor with direct bandgap is expected to have a constant density of states (DOS) at the band edges within a quasi-particle picture \cite{grundmann_physics_2016}. 
The optical response, which is directly related to the joint DOS, should ideally have a linear lineshape due to the convolution of two constant functions. 
However, in most practical TMDC systems, the insufficient screening generally enhances the Coulomb interaction between the electrons and holes, leading to exciton formation and complicating the lineshapes. 
Intriguingly, by placing the TMDC on a metal surface, it may be possible to recover the intrinsic exciton-free optical response of the 2D material. 
In this case, it is possible to extract the onset of the band edge from the linear extrapolation, analogous to the \textit{Tauc plot}.
Regrettably, when the TMDC is positioned on a metal surface, it is extremely difficult to isolate its pure optical response from that of the underlying metallic substrate. 
For instance, in the case of a \ce{MoS2} monolayer on Au, the photoluminescence (PL) signal is completely quenched \cite{bhanu_photoluminescence_2015,pollmann_large-area_2021}. 
While absorption or reflectivity measurements can still provide an optical response from the TMDC monolayer on a metal system, the extensive optical response from the free electron of the substrate often obscures minor exciton features \cite{zou_spectroscopic_2021}, even though some subtle features in the first derivative of the spectrum have been reported \cite{park_direct_2018}. 
These persistent challenges impede a comprehensive understanding of the optical behavior of TMDC monolayers interfaced with a metal substrate.

Sum frequency generation spectroscopy, a second-order nonlinear spectroscopic technique\cite{shen_surface_1989,boyd_nonlinear_2020}, offers sensitivity to the structural symmetry of \ce{MoS2}. 
In a prior study, we demonstrated the possibility of isolating the optical response of \ce{MoS2} from that of the metal by leveraging their respective symmetries \cite{yang_interaction_2023}. 
In this work, we extensively investigate the influence of a metallic substrate on the spectra of \ce{MoS2}. 
For the first time, a linear optical lineshape of the TMDC monolayer has been recovered at the band edge. 
This not only allows us to unambiguously conclude that the commonly observed A and B excitons disappear under such extreme screening and doping conditions, but also provides previously unrevealed information to discuss the unique electronic property of the TMDC monolayer in contact with metal with the help of density functional theory (DFT) calculations. 
Given the disappearance of the exciton features, we can derive an analogous \textit{Tauc plot} method using the pure second-order nonlinear susceptibility ($\chi^{(2)}$) of \ce{MoS2}. 
With this method, we succeed in extracting the intrinsic bandgap ($1.65\pm0.20$ eV) of \ce{MoS2} within the semiconductor-metal junction. 
Many interesting properties of the system are discussed, such as bandgap renormalization, quantum confinement, and light-matter interaction. 
This study can be extended to other 2D semiconductor and 3D metal junctions which are commonly encountered in optoelectronic and photocatalytic devices. The information obtained can greatly improve our understanding of the optical properties of 2D semiconductors, especially in the presence of significant concentrations of free carriers.

\section{Results and Discussion}

\begin{figure}[t]
    \centering
    \includegraphics[scale=0.4]{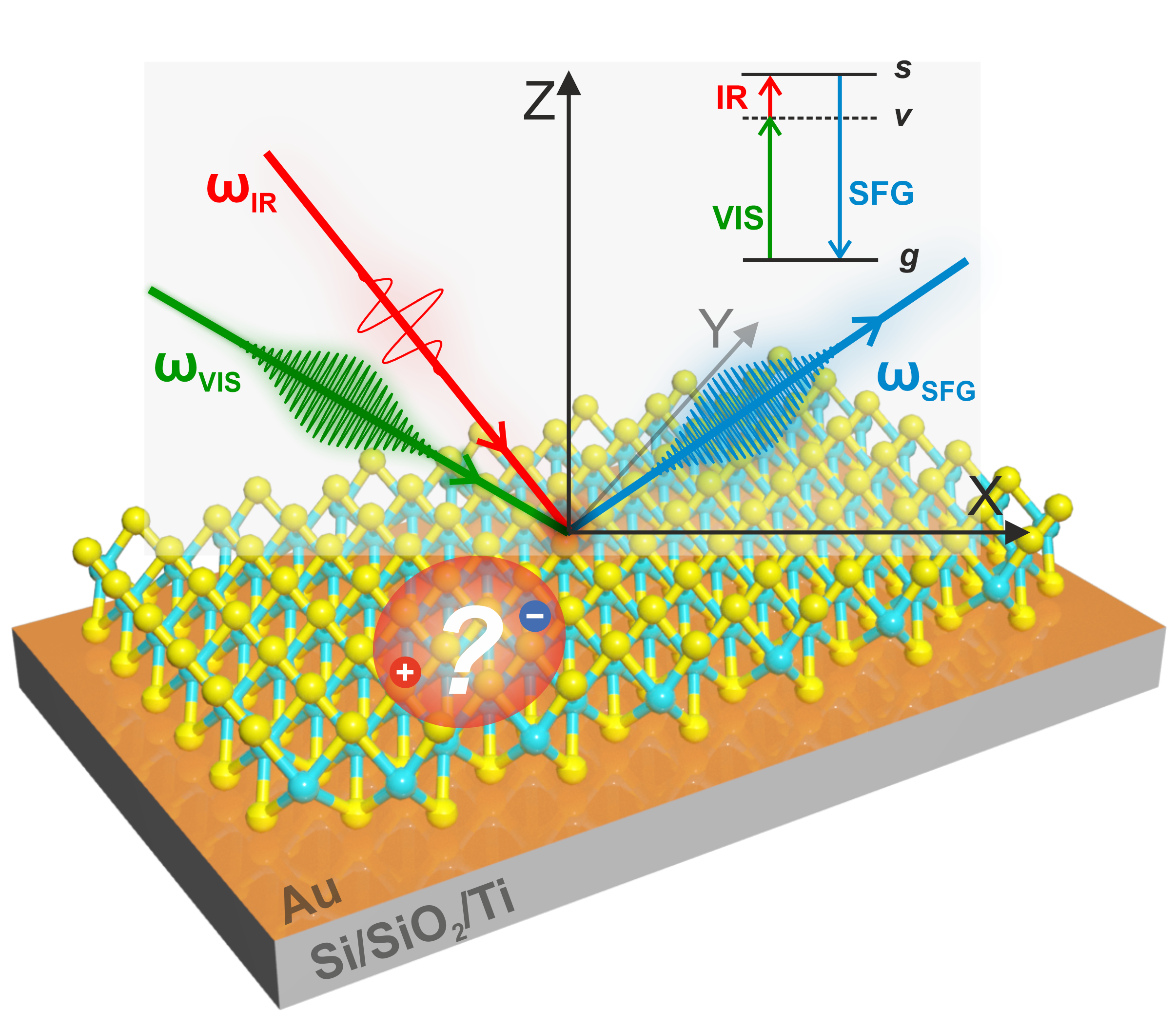}
    \caption{Schematic representations of exfoliated \ce{MoS2} monolayer on Au is investigated by the FS-SFG. Insert: Schematic energy level diagram of the FS-SFG, where g, v, and s represent the ground state, virtual state, and excited state, respectively. Only the SFG process will resonate with the exciton state in \ce{MoS2} monolayer.}
    \label{fig:setup}
\end{figure}

In this study, we collected final-state sum frequency generation (FS-SFG) signals from a monolayer of \ce{MoS2} on Au, where the Au film was deposited via physical vapor deposition (PVD). 
The \ce{MoS2}/Au sample was prepared through mechanical exfoliation of \ce{MoS2} onto a freshly prepared Au surface \cite{pollmann_large-area_2021}. 
For comparative analysis, we also investigated a monolayer of \ce{MoS2} on \ce{SiO2}. Optical images of \ce{MoS2}/Au and \ce{MoS2}/\ce{SiO2} can be referenced in Fig. S1 within the Supplementary Material  \cite{[{See Supplemental Material at [\textbf{URL will be inserted by publisher}] for experimental and theoretical information, which includes the optical images of the samples,  normalized SFG spectra of PVD Au and \ce{MoS2}/\ce{SiO2}, details about sample preparation and laser setup, DFT-calculated band structure and density of states of \ce{MoS2} monolayer on Au. Furthermore, justification of using $|\chi^{(2)}|$ to characterize the band edge}] Yang_spectrum_2023}. 
A detailed depiction of the SFG experimental setup is provided in Fig. \ref{fig:setup}. 
Two laser beams were spatially and temporally overlapped on the sample surface, and the resulting emitted sum frequency photons were detected. 
The visible beam's photon energy was centered at 1.56 eV, while the photon energy of infrared (IR) beam was tuned from 0.28-0.41 eV, ensuring that the final-state resonant SFG photon energies covered the A and B exciton energies. 
The visible beam energy was set at 0.8 $\mu$J/pulse, and the infrared beam was maintained below 1.0 $\mu$J/pulse. Both beams were propagated coplanarly with incident angles of 64$^{\circ}$ and 46$^{\circ}$ for the visible and infrared beams, respectively. 
The polarization of the three beams was individually set to \textit{p} or \textit{s} to obtain different polarization combinations, allowing us to choose the one best suited for the intended experiment. 
During measurements, azimuthal-dependent patterns were initially obtained to guide the collection of FS-SFG spectra at specific azimuthal angles, e.g. the angle with maximum intensity. 
A $z$-cut alpha quartz was used to obtain a reference signal for the purpose of correcting the frequency-dependent IR intensity. All experiments were conducted under ambient conditions at $\sim21.5^{\circ}$C. 
Further details regarding the sample preparation and characterization and the laser setup can be found in our previous work  \cite{pollmann_large-area_2021,tong_hydrophobic_2017,tong_experimentally_2018} and Supplemental Material \cite{Yang_spectrum_2023}.

\begin{figure}[t]
   \centering
   \includegraphics[scale=1.2]{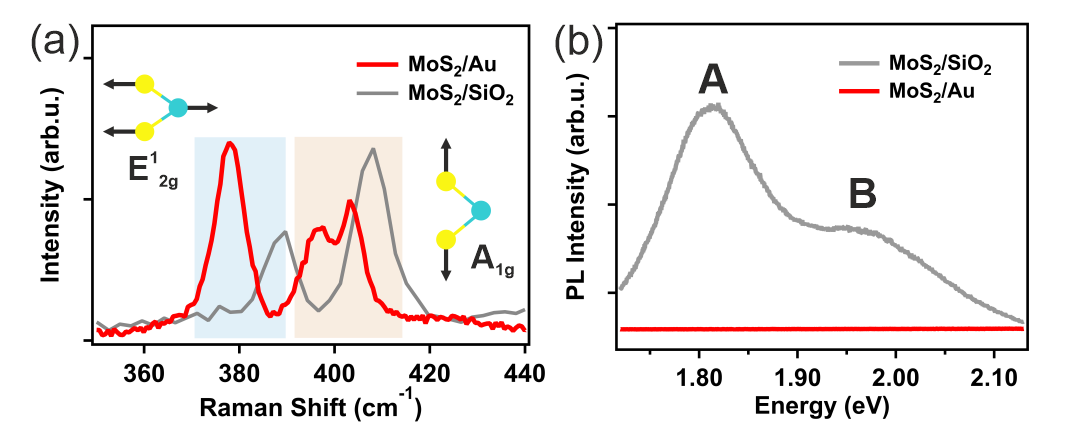}
    \caption{(a) Raman spectra of \ce{MoS2} monolayer on Au and Si/\ce{SiO2} substrate. The inset illustrates two Raman-active modes ($E^1_{2g}$ and $A_{1g}$). (b) PL spectra obtained at room temperature for \ce{MoS2} monolayer on Au and Si/\ce{SiO2} substrate with an excitation wavelength of 532 nm. The PL signal of \ce{MoS2} on Au is strongly quenched.}
    \label{fig:Raman}
\end{figure}

To gain a basic overview of our samples, we initiated our study with Raman and PL measurements. As depicted in Fig. \ref{fig:Raman}, both Raman and PL spectra of \ce{MoS2} monolayers on Au and \ce{SiO2} substrates manifest pronounced distinctions. Specifically, the Raman spectrum of the \ce{MoS2} monolayer on \ce{SiO2}, as illustrated in Fig. \ref{fig:Raman}(a), exhibits two notable peaks centered at 388.5 cm$^{-1}$ for the out-of-plane $E^1_{2g}$ mode and 407.8 cm$^{-1}$ for the in-plane $A_{1g}$ mode, respectively. The discernible 19.3 cm$^{-1}$ difference between these modes is characteristic of a monolayer structure \cite{lee_anomalous_2010}. However, the Raman spectrum of \ce{MoS2} on Au showcases markedly disparate features. The $E^1_{2g}$ mode undergoes a redshift to 377.9 cm$^{-1}$, and the $A_{1g}$ mode exhibits a red shift and splits into two distinct peaks at 396.2 and 403.6 cm$^{-1}$. These notable changes align with earlier observations \cite{velicky_strain_2020,pollmann_large-area_2021} and can be attributed to the considerable tensile strain induced by the Au substrate owing to the high-quality contact \cite{pollmann_large-area_2021, sarkar_signatures_2021}.

The substantial interaction observed at the semiconductor-metal interface is further corroborated by the PL spectrum depicted in Fig. \ref{fig:Raman}(b). 
The \ce{MoS2} sample on \ce{SiO2} exhibits distinctive features at 1.81 and 1.97 eV, corresponding to the A and B excitons of \ce{MoS2}, respectively. 
The presence of two excitons, A and B, originates from the splitting of the valence band maxima at the K-point, induced by spin-orbit coupling \cite{xiao_coupled_2012}. 
In stark contrast, the PL signal from the \ce{MoS2}/Au sample is completely quenched. These optical spectra observations align with previous studies \cite{scheuschner_photoluminescence_2014,velicky_strain_2020,pollmann_large-area_2021}. 
PL quenching has been reported previously \cite{bhanu_photoluminescence_2015} and was understood as the transfer of the photoexcited electron from the conduction band of \ce{MoS2} to the Au substrate instead of undergoing radiation relaxation.
This hypothesis is supported by tr-ARPES results, which demonstrate ultrafast dynamics of the photoexcited free carriers \cite{cabo_observation_2015}. 
However, the behavior of excitons within this heterointerface remained unclear. 
While there are reports claiming that the A and B excitons remain observable for \ce{MoS2} on Au \cite{park_direct_2018}, others suggest a conclusion to the contrary \cite{zou_spectroscopic_2021}. 
The reason for the dispute is the weak signal of \ce{MoS2}, which is always hidden under the strong background signal of the Au substrate, making it difficult to unambiguously distinguish between them. 
Fortunately, second-order nonlinear spectroscopy techniques, such as FS-SFG, provide a unique opportunity to isolate the contribution of the monolayer semiconductor from that of the bulk substrate, capitalizing on their distinct symmetries. Further details regarding this isolation technique are elaborated below.

As demonstrated in earlier studies utilizing single-beam second harmonic generation (SHG), the 2H-\ce{MoS2} monolayer is categorized under the $D_{3h}$ point group, displaying a consistent six-fold symmetric pattern of SHG intensity \cite{kumar_second_2013,malard_observation_2013,jiang_valley_2014}. 
Our recent investigation using SFG aligns with these prior findings and extends the understanding by revealing that the symmetric pattern of the \ce{MoS2} monolayer on Au is highly dependent on polarization, resulting in a lowering of the symmetry to $C_{3v}$ \cite{yang_interaction_2023}. 
Within the $C_{3v}$ point group, specific polarization combinations (e.g. \textit{spp}, \textit{pps}, \textit{psp}, and \textit{sss}, where \textit{spp} indicates \textit{s} polarized SFG, \textit{p} polarized visible, and \textit{p} polarized infrared beams) yield signals arising from only one azimuthal-dependent second-order susceptibility component inherited from \ce{MoS2}. 
By carefully selecting the appropriate polarization combination, SFG enables selective investigation of the \ce{MoS2} monolayer and provides the pure response from the \ce{MoS2} monolayer free from perturbations by the Au optical response.

\begin{figure}[t]
    \centering
    \includegraphics[scale=1]{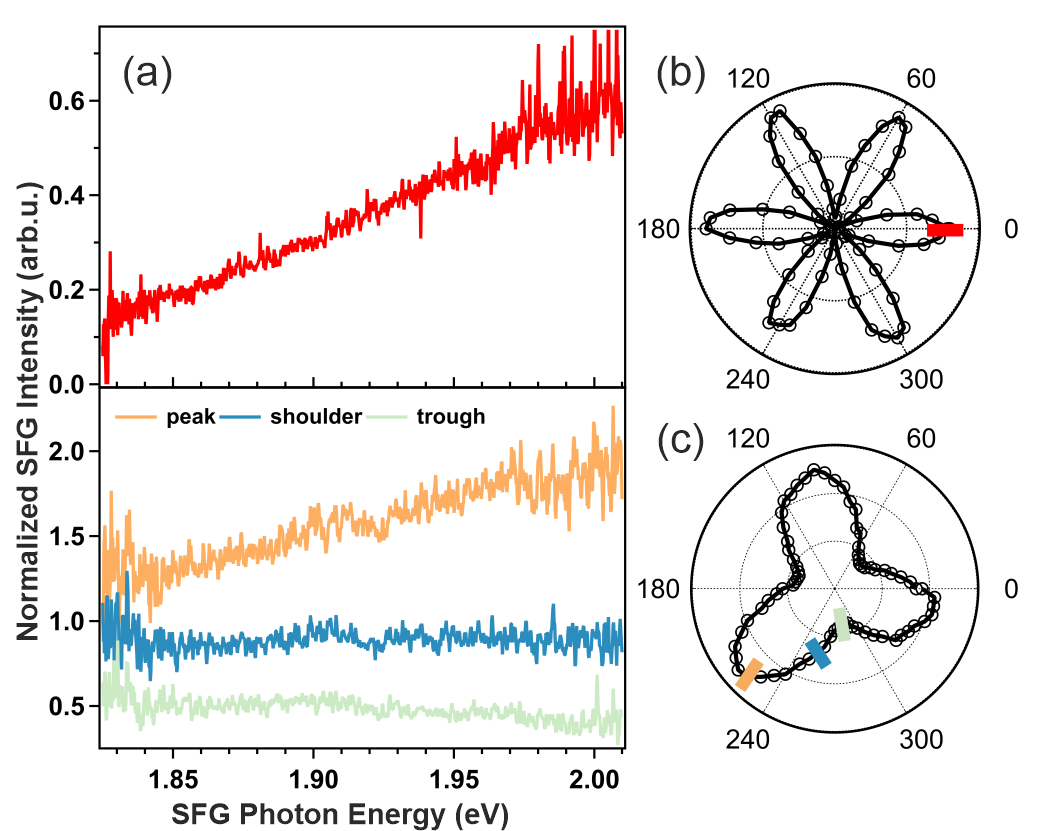}
    \caption{(a) The normalized FS-SFG spectra of \ce{MoS2} monolayer on Au under \textit{spp} (top) and \textit{ssp} (bottom) polarization combinations. (b) and (c) are the azimuthal-dependent SFG intensity of \ce{MoS2}/Au under \textit{spp} and \textit{ssp} polarization combination, respectively. There is a $13^{\circ}$ offset in the azimuthal angle between \textit{spp} and \textit{ssp} polarization combination. The spectra in (a) were collected at specific azimuthal angles, marked in (b) and (c) with corresponding colors.}
    \label{fig:SFG}
\end{figure}

Figure \ref{fig:SFG}(a) top and bottom panels show representative FS-SFG spectra of the \ce{MoS2}/Au sample under \textit{spp} and \textit{ssp} polarization combinations.  
Figure \ref{fig:SFG}(b) and (c) are the corresponding azimuthal dependencies of the SFG intensity for the two polarization combinations. 
The azimuthal pattern for \textit{spp} exhibits a six-fold symmetry, while that for \textit{ssp} shows a three-fold symmetry. 
The reason for the different azimuthal symmetric pattern has been extensively discussed in our previous work \cite{yang_interaction_2023}. 
In brief, for \textit{spp} polarization, the SFG signal purely originates from the \ce{MoS2}, while on the other hand both \ce{MoS2} and Au contribute to the SFG signal when using \textit{ssp} polarization. 
In the latter case, the interference of the two signals results in a reduction in the symmetry from six-fold to three-fold. 

In this work, we will focus on the spectra of these FS-SFG responses. 
For \textit{spp} polarization combination, the spectral shape in the reported A and B exciton energy region is close to a straight line and the slope is independent of the azimuthal angle, only the intensities change (Here only a representative spectrum collected at the maximum integrated FS-SFG intensities, which is marked in red in Fig. \ref{fig:SFG}(b) is presented).
On the contrary, the spectral shape recorded under the \textit{ssp} polarization combination (bottom panel of Fig. \ref{fig:SFG}(a)) maintains a linear appearance, yet with slopes that exhibit significant sensitivity to the azimuthal angle (the weak narrow features around 1.92 eV are from the vibrational responses of the unavoidable hydrocarbon impurities being upconverted by the 1.56 eV visible beam). Specifically, at the azimuthal intensity maximum (highlighted in orange in Fig. \ref{fig:SFG}(c)), the slope is positive; at the azimuthal intensity minimum (marked in light-green in Fig. \ref{fig:SFG}(c)), the slope is negative. Furthermore, at the shoulder, the spectrum appears nearly horizontal (indicated by the blue marker in Fig. \ref{fig:SFG}(c)), akin to the FS-SFG response observed in pure PVD Au (refer to Fig. S2 in the Supplemental Material \cite{Yang_spectrum_2023}).
To establish a reference, azimuthal-dependent SFG intensity and spectrum of \ce{MoS2}/\ce{SiO2} were also obtained.  
A distinctive resonant peak is observed at $\sim$1.84 eV, attributed to the A exciton, as illustrated in Fig. S3 of the Supplemental Material \cite{Yang_spectrum_2023}. 
Clearly, in the case of \ce{MoS2}/Au, even though the FS-SFG photon energy falls within the A and B exciton energy range, all the spectra display a linear profile, and no excitonic features were observed.  

While both the PL (Fig. \ref{fig:Raman}) and the FS-SFG (Fig. \ref{fig:SFG}) spectra suggest a quench of the excitonic features, the implications of the two observations differ significantly. 
The quenched spectrum in PL solely implies a non-radiative relaxation process, and it does not provide any additional information such as bandgap renormalization or potential substrate induced doping.
On the contrary, the FS-SFG spectra of \ce{MoS2} on Au distinctly exhibit the radiative signal. A comprehensive analysis of this signal can provide valuable insights into the electronic properties of this intricate system.

One of the intriguing properties of 2D materials is their extreme sensitivity to the screening from the environment. Both theory \cite{olsen_simple_2016,druppel_diversity_2017,steinhoff_frequency-dependent_2018} and experiments \cite{ugeda_giant_2014,cabo_observation_2015} have shown that the screening from the environment or from the substrate doped free charge carriers inside the TMDC can effectively reduce the Coulomb interaction between electron-hole pairs, leading to an increase in their spatial separation and a reduction in the exciton's binding energy. Depending on the nature of screening (dynamic or static), the quasiparticle bandgap may be reduced (known as bandgap renormalization) by the same amount as the change of the binding energy of the exciton \cite{druppel_diversity_2017}.   
Obviously, metals possess an extremely large dielectric constant and produce a strong doping effect compared to free-standing or dielectric substrate-supported conditions. Consequently, stable excitons cannot be formed in a \ce{MoS2} monolayer on Au at room temperature, as evidenced by our observation.  

\begin{figure}[t]
    \centering
    \includegraphics[scale=1]{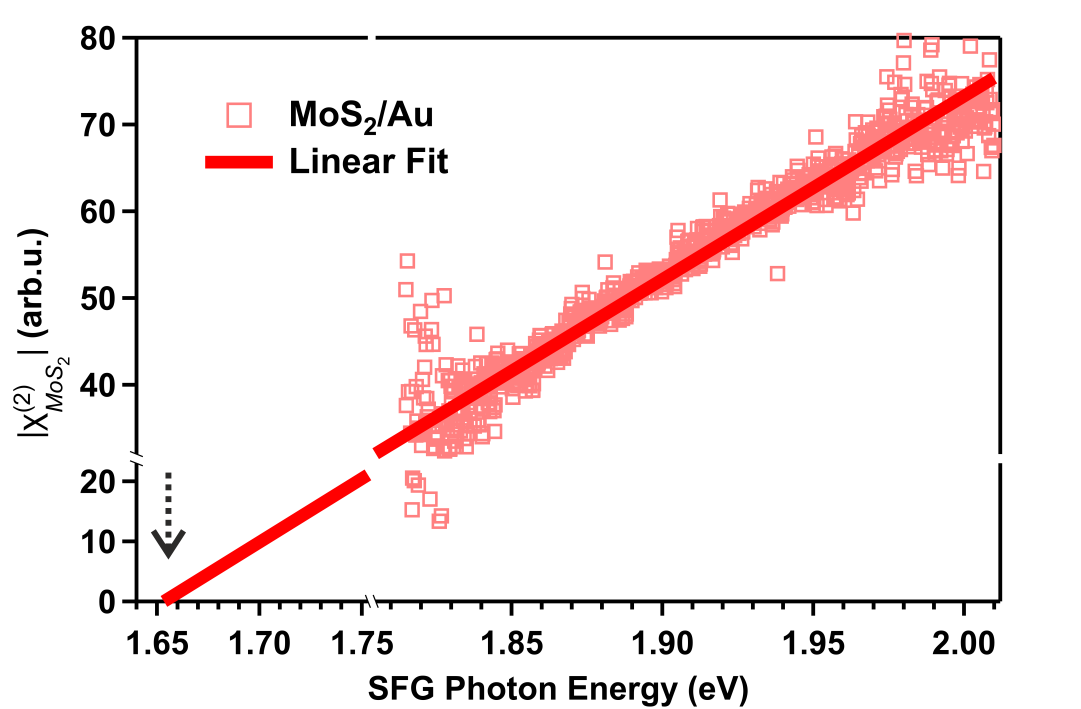}
    \caption{The onset of bandgap in \ce{MoS2} monolayer on Au under \textit{spp} polarization combination was determined using the analogous \textit{Tauc plot} method. A linear fit of the dependence of $|\chi_{MoS_2}^{(2)}|$ on the SFG photon energy was represented by the red solid line. The onset of the bandgap was extracted by extrapolating the linear fit to hit the energy axis ($|\chi_{MoS_2}^{(2)}|=0$) and marked by an arrow.}
    \label{fig:Tauc}
\end{figure}

As the formation of excitons within \ce{MoS2}/Au is not feasible, the isolated optical response of the \ce{MoS2} hence can report the electronic property of the semiconductor at the band edge, similar to the \textit{Tauc plot} method reported for bulk amorphous semiconductors \cite{klein_limitations_2023}. 
Fig. \ref{fig:Tauc} shows the results of such an analogous \textit{Tauc plot}. A straight line is clearly obtained. 
Extrapolating the line in the plot to $|\chi_{MoS_2}^{(2)}|=0$, as shown in Fig. \ref{fig:Tauc}, gives a value of $1.65\pm0.20$ eV as averaged from several independent measurements. 
Given that this response is approximately equivalent to the convolution of two constant DOS functions at the band edges (as confirmed by a DFT calculation), this value reports the onset of the bandgap at the K-point in the first Brillouin zone (see Supplemental Material \cite{Yang_spectrum_2023} about the band structure and constant DOS). 

The \textit{Tauc plot} method has previously been utilized to determine the bandgap of TMDC thin films, without accounting for the impact of excitons \cite{zheng_photoconductivity_2020,paul_optical_2022}. However, for a free-standing TMDC monolayer or a monolayer on a dielectric substrate, the presence of excitons due to insufficient screening, can significantly hinder the unambiguous identification of the absorption edge. 
In contrast, in the current study, the exciton binding energy is sufficiently low when the TMDC monolayer is placed on Au, so that the analogous \textit{Tauc plot} becomes possible as long as the pure monolayer response can be extracted. 
It should be noted that in this study we directly plot the Fresnel factor corrected $|\chi_{MoS_2}^{(2)}|$ as a function of the sum frequency photon energy, rather than plotting $|\alpha_{MoS_2}h\nu|^n$ as a function of the input photon energy $h\nu$. This approach is justified by the Miller theory in nonlinear optics \cite{boyd_nonlinear_2020}, where the lineshapes of a final state resonance in a second-order nonlinear optical process are identical to those of a linear optical process such as optical absorption (see Supplemental Material \cite{Yang_spectrum_2023} for further discussion).    

The quasiparticle bandgap obtained in this study is much smaller than that for free-standing or dielectric material supported \ce{MoS2} monolayer. The screening effect from the substrate and from the additional charge carriers inside the monolayer via substrate doping may be the main reason. 
A tr-ARPES study revealed a direct electronic bandgap of 1.95 eV at the K-point \cite{cabo_observation_2015}.
However, it should be noted that the bandgap measured in their study refers to the energy difference between the center of the lowest conduction band and that of the highest valence bands at the K-point. Whereas the value extracted from our experiment is the onset of the difference between these two. Considering the full width at the half maximum ($\sim$0.3 eV) of the bands in the tr-APRES results (Fig. 2 in ref \cite{cabo_observation_2015}), our results are in agreement with theirs.  
In a STS study, a bandgap of $1.74\pm0.27$ eV was extracted from the STS spectra \cite{bruix_single-layer_2016}, aligning closely with our result. 
Additionally, different quasiparticle bandgap values, ranging from 1.6 to 2.9 eV, have been reported from previous calculations using different level of theories \cite{cheiwchanchamnangij_quasiparticle_2012,kormanyos_kp_2015,druppel_diversity_2017}. Our calculation provides new and reliable parameters for future comparison between the theory and experiment. 

In addition to the electronic bandgap, the interfaces created when 2D semiconductors come into contact with 3D metallic electrodes play a pivotal role in many newly developed optoelectronic and photocatalytic devices \cite{wang_van_2019,shen_ultralow_2021}. 
Our study offers a straightforward optical method to characterize the optical response of the 2D semiconductor itself with a non-contact manner. 
Such a method can be extended to other similar systems for devices based on the optoelectronic properties of TMDCs.  

While our current research is both novel and informative, it is important to acknowledge certain limitations. Specifically, the limited probing frequency range is a result of constraints imposed by our experimental setup. Additionally, in the determination of the bandgap onset using the analogous \textit{Tauc plot}, we had to employ an approximation based on Miller's rule \cite{boyd_nonlinear_2020}. However, despite these constraints, our conclusions regarding the absence of A and B excitons and the observed reduction of the bandgap for \ce{MoS2}/Au are firmly rooted in the experimental data. To address these limitations and deepen our understanding, we plan to widen the optical window and conduct heterodyne SFG measurements in our future studies.

\section{Conclusion}
Through careful polarization selection in the FS-SFG measurements, we were able to effectively isolate the \ce{MoS2} monolayer's optical response from the \ce{MoS2}/Au system. 
The significant screening effect induced by the Au substrate and by the doped free carriers inside the monolayer results in a substantial reduction of the exciton's binding energy and remarkable bandgap renormalization, effectively impeding the formation of A and B excitons.
Under these conditions, we could extract the expected linear onset of optical absorption of a TMDC semiconductor under a quasi-particle picture. Our study provides valuable insights for understanding the electronic properties at the 2D semiconductor and 3D metal junction with the help of DFT calculations. By introducing an analogous \textit{Tauc plot} using the pure second-order nonlinear susceptibility ($|\chi_{MoS_2}^{(2)}|$), we accurately determine a quasiparticle bandgap of $1.65\pm0.20$ eV. This methodology can be applied to various 2D semiconductor and 3D metal junctions, improving our understanding of optical properties in the presence of abundant free carriers and providing a valuable reference for the further development of photocatalysts.

\begin{acknowledgments}
This work was funded by the Deutsche Forschungsgemeinschaft (DFG, German Research Foundation) through projects A06, B02, and C05 within the SFB 1242 "Non-Equilibrium Dynamics of Condensed Matter in the Time Domain" (project number 278162697), through Germany's Excellence Strategy (EXC 2033 - 390677874 - RESOLV). Additional support was provided by the European Research Council, i.e., ERC- CoG-2017 SOLWET (project number 772286) to RKC. 

\end{acknowledgments}



\begin{thebibliography}{50}%
	\makeatletter
	\providecommand \@ifxundefined [1]{%
		\@ifx{#1\undefined}
	}%
	\providecommand \@ifnum [1]{%
		\ifnum #1\expandafter \@firstoftwo
		\else \expandafter \@secondoftwo
		\fi
	}%
	\providecommand \@ifx [1]{%
		\ifx #1\expandafter \@firstoftwo
		\else \expandafter \@secondoftwo
		\fi
	}%
	\providecommand \natexlab [1]{#1}%
	\providecommand \enquote  [1]{``#1''}%
	\providecommand \bibnamefont  [1]{#1}%
	\providecommand \bibfnamefont [1]{#1}%
	\providecommand \citenamefont [1]{#1}%
	\providecommand \href@noop [0]{\@secondoftwo}%
	\providecommand \href [0]{\begingroup \@sanitize@url \@href}%
	\providecommand \@href[1]{\@@startlink{#1}\@@href}%
	\providecommand \@@href[1]{\endgroup#1\@@endlink}%
	\providecommand \@sanitize@url [0]{\catcode `\\12\catcode `\$12\catcode
		`\&12\catcode `\#12\catcode `\^12\catcode `\_12\catcode `\%12\relax}%
	\providecommand \@@startlink[1]{}%
	\providecommand \@@endlink[0]{}%
	\providecommand \url  [0]{\begingroup\@sanitize@url \@url }%
	\providecommand \@url [1]{\endgroup\@href {#1}{\urlprefix }}%
	\providecommand \urlprefix  [0]{URL }%
	\providecommand \Eprint [0]{\href }%
	\providecommand \doibase [0]{https://doi.org/}%
	\providecommand \selectlanguage [0]{\@gobble}%
	\providecommand \bibinfo  [0]{\@secondoftwo}%
	\providecommand \bibfield  [0]{\@secondoftwo}%
	\providecommand \translation [1]{[#1]}%
	\providecommand \BibitemOpen [0]{}%
	\providecommand \bibitemStop [0]{}%
	\providecommand \bibitemNoStop [0]{.\EOS\space}%
	\providecommand \EOS [0]{\spacefactor3000\relax}%
	\providecommand \BibitemShut  [1]{\csname bibitem#1\endcsname}%
	\let\auto@bib@innerbib\@empty
	\bibitem [{\citenamefont {Mak}\ \emph {et~al.}(2010)\citenamefont {Mak},
		\citenamefont {Lee}, \citenamefont {Hone}, \citenamefont {Shan},\ and\
		\citenamefont {Heinz}}]{mak_atomically_2010}%
	\BibitemOpen
	\bibfield  {author} {\bibinfo {author} {\bibfnamefont {K.~F.}\ \bibnamefont
			{Mak}}, \bibinfo {author} {\bibfnamefont {C.}~\bibnamefont {Lee}}, \bibinfo
		{author} {\bibfnamefont {J.}~\bibnamefont {Hone}}, \bibinfo {author}
		{\bibfnamefont {J.}~\bibnamefont {Shan}},\ and\ \bibinfo {author}
		{\bibfnamefont {T.~F.}\ \bibnamefont {Heinz}},\ }\href
	{https://doi.org/10.1103/PhysRevLett.105.136805} {\bibfield  {journal}
		{\bibinfo  {journal} {Phys. Rev. Lett.}\ }\textbf {\bibinfo {volume} {105}},\
		\bibinfo {pages} {136805} (\bibinfo {year} {2010})}\BibitemShut {NoStop}%
	\bibitem [{\citenamefont {Splendiani}\ \emph {et~al.}(2010)\citenamefont
		{Splendiani}, \citenamefont {Sun}, \citenamefont {Zhang}, \citenamefont {Li},
		\citenamefont {Kim}, \citenamefont {Chim}, \citenamefont {Galli},\ and\
		\citenamefont {Wang}}]{splendiani_emerging_2010}%
	\BibitemOpen
	\bibfield  {author} {\bibinfo {author} {\bibfnamefont {A.}~\bibnamefont
			{Splendiani}}, \bibinfo {author} {\bibfnamefont {L.}~\bibnamefont {Sun}},
		\bibinfo {author} {\bibfnamefont {Y.}~\bibnamefont {Zhang}}, \bibinfo
		{author} {\bibfnamefont {T.}~\bibnamefont {Li}}, \bibinfo {author}
		{\bibfnamefont {J.}~\bibnamefont {Kim}}, \bibinfo {author} {\bibfnamefont
			{C.-Y.}\ \bibnamefont {Chim}}, \bibinfo {author} {\bibfnamefont
			{G.}~\bibnamefont {Galli}},\ and\ \bibinfo {author} {\bibfnamefont
			{F.}~\bibnamefont {Wang}},\ }\href {https://doi.org/10.1021/nl903868w}
	{\bibfield  {journal} {\bibinfo  {journal} {Nano Lett.}\ }\textbf {\bibinfo
			{volume} {10}},\ \bibinfo {pages} {1271} (\bibinfo {year}
		{2010})}\BibitemShut {NoStop}%
	\bibitem [{\citenamefont {Wang}\ \emph {et~al.}(2012)\citenamefont {Wang},
		\citenamefont {Kalantar-Zadeh}, \citenamefont {Kis}, \citenamefont
		{Coleman},\ and\ \citenamefont {Strano}}]{wang_electronics_2012}%
	\BibitemOpen
	\bibfield  {author} {\bibinfo {author} {\bibfnamefont {Q.~H.}\ \bibnamefont
			{Wang}}, \bibinfo {author} {\bibfnamefont {K.}~\bibnamefont
			{Kalantar-Zadeh}}, \bibinfo {author} {\bibfnamefont {A.}~\bibnamefont {Kis}},
		\bibinfo {author} {\bibfnamefont {J.~N.}\ \bibnamefont {Coleman}},\ and\
		\bibinfo {author} {\bibfnamefont {M.~S.}\ \bibnamefont {Strano}},\ }\href
	{https://doi.org/10.1038/nnano.2012.193} {\bibfield  {journal} {\bibinfo
			{journal} {Nat. Nanotechnol.}\ }\textbf {\bibinfo {volume} {7}},\ \bibinfo
		{pages} {699} (\bibinfo {year} {2012})}\BibitemShut {NoStop}%
	\bibitem [{\citenamefont {Mak}\ and\ \citenamefont
		{Shan}(2016)}]{mak_photonics_2016}%
	\BibitemOpen
	\bibfield  {author} {\bibinfo {author} {\bibfnamefont {K.~F.}\ \bibnamefont
			{Mak}}\ and\ \bibinfo {author} {\bibfnamefont {J.}~\bibnamefont {Shan}},\
	}\href {https://doi.org/10.1038/nphoton.2015.282} {\bibfield  {journal}
		{\bibinfo  {journal} {Nat. Photon.}\ }\textbf {\bibinfo {volume} {10}},\
		\bibinfo {pages} {216} (\bibinfo {year} {2016})}\BibitemShut {NoStop}%
	\bibitem [{\citenamefont {Mueller}\ and\ \citenamefont
		{Malic}(2018)}]{mueller_exciton_2018}%
	\BibitemOpen
	\bibfield  {author} {\bibinfo {author} {\bibfnamefont {T.}~\bibnamefont
			{Mueller}}\ and\ \bibinfo {author} {\bibfnamefont {E.}~\bibnamefont
			{Malic}},\ }\href {https://doi.org/10.1038/s41699-018-0074-2} {\bibfield
		{journal} {\bibinfo  {journal} {npj 2D Mater Appl}\ }\textbf {\bibinfo
			{volume} {2}},\ \bibinfo {pages} {1} (\bibinfo {year} {2018})}\BibitemShut
	{NoStop}%
	\bibitem [{\citenamefont {Liu}\ \emph {et~al.}(2021)\citenamefont {Liu},
		\citenamefont {Duan}, \citenamefont {Shin}, \citenamefont {Park},
		\citenamefont {Huang},\ and\ \citenamefont {Duan}}]{liu_promises_2021}%
	\BibitemOpen
	\bibfield  {author} {\bibinfo {author} {\bibfnamefont {Y.}~\bibnamefont
			{Liu}}, \bibinfo {author} {\bibfnamefont {X.}~\bibnamefont {Duan}}, \bibinfo
		{author} {\bibfnamefont {H.-J.}\ \bibnamefont {Shin}}, \bibinfo {author}
		{\bibfnamefont {S.}~\bibnamefont {Park}}, \bibinfo {author} {\bibfnamefont
			{Y.}~\bibnamefont {Huang}},\ and\ \bibinfo {author} {\bibfnamefont
			{X.}~\bibnamefont {Duan}},\ }\href
	{https://doi.org/10.1038/s41586-021-03339-z} {\bibfield  {journal} {\bibinfo
			{journal} {Nature}\ }\textbf {\bibinfo {volume} {591}},\ \bibinfo {pages}
		{43} (\bibinfo {year} {2021})}\BibitemShut {NoStop}%
	\bibitem [{\citenamefont {Radisavljevic}\ \emph {et~al.}(2011)\citenamefont
		{Radisavljevic}, \citenamefont {Radenovic}, \citenamefont {Brivio},
		\citenamefont {Giacometti},\ and\ \citenamefont
		{Kis}}]{radisavljevic_single-layer_2011}%
	\BibitemOpen
	\bibfield  {author} {\bibinfo {author} {\bibfnamefont {B.}~\bibnamefont
			{Radisavljevic}}, \bibinfo {author} {\bibfnamefont {A.}~\bibnamefont
			{Radenovic}}, \bibinfo {author} {\bibfnamefont {J.}~\bibnamefont {Brivio}},
		\bibinfo {author} {\bibfnamefont {V.}~\bibnamefont {Giacometti}},\ and\
		\bibinfo {author} {\bibfnamefont {A.}~\bibnamefont {Kis}},\ }\href
	{https://doi.org/10.1038/nnano.2010.279} {\bibfield  {journal} {\bibinfo
			{journal} {Nat. Nanotechnol.}\ }\textbf {\bibinfo {volume} {6}},\ \bibinfo
		{pages} {147} (\bibinfo {year} {2011})}\BibitemShut {NoStop}%
	\bibitem [{\citenamefont {Yin}\ \emph {et~al.}(2012)\citenamefont {Yin},
		\citenamefont {Li}, \citenamefont {Li}, \citenamefont {Jiang}, \citenamefont
		{Shi}, \citenamefont {Sun}, \citenamefont {Lu}, \citenamefont {Zhang},
		\citenamefont {Chen},\ and\ \citenamefont {Zhang}}]{yin_single-layer_2012}%
	\BibitemOpen
	\bibfield  {author} {\bibinfo {author} {\bibfnamefont {Z.}~\bibnamefont
			{Yin}}, \bibinfo {author} {\bibfnamefont {H.}~\bibnamefont {Li}}, \bibinfo
		{author} {\bibfnamefont {H.}~\bibnamefont {Li}}, \bibinfo {author}
		{\bibfnamefont {L.}~\bibnamefont {Jiang}}, \bibinfo {author} {\bibfnamefont
			{Y.}~\bibnamefont {Shi}}, \bibinfo {author} {\bibfnamefont {Y.}~\bibnamefont
			{Sun}}, \bibinfo {author} {\bibfnamefont {G.}~\bibnamefont {Lu}}, \bibinfo
		{author} {\bibfnamefont {Q.}~\bibnamefont {Zhang}}, \bibinfo {author}
		{\bibfnamefont {X.}~\bibnamefont {Chen}},\ and\ \bibinfo {author}
		{\bibfnamefont {H.}~\bibnamefont {Zhang}},\ }\href
	{https://doi.org/10.1021/nn2024557} {\bibfield  {journal} {\bibinfo
			{journal} {ACS Nano}\ }\textbf {\bibinfo {volume} {6}},\ \bibinfo {pages}
		{74} (\bibinfo {year} {2012})}\BibitemShut {NoStop}%
	\bibitem [{\citenamefont {Bartolomeo}\ \emph {et~al.}(2017)\citenamefont
		{Bartolomeo}, \citenamefont {Genovese}, \citenamefont {Giubileo},
		\citenamefont {Iemmo}, \citenamefont {Luongo}, \citenamefont {Foller},\ and\
		\citenamefont {Schleberger}}]{bartolomeo_hysteresis_2017}%
	\BibitemOpen
	\bibfield  {author} {\bibinfo {author} {\bibfnamefont {A.~D.}\ \bibnamefont
			{Bartolomeo}}, \bibinfo {author} {\bibfnamefont {L.}~\bibnamefont
			{Genovese}}, \bibinfo {author} {\bibfnamefont {F.}~\bibnamefont {Giubileo}},
		\bibinfo {author} {\bibfnamefont {L.}~\bibnamefont {Iemmo}}, \bibinfo
		{author} {\bibfnamefont {G.}~\bibnamefont {Luongo}}, \bibinfo {author}
		{\bibfnamefont {T.}~\bibnamefont {Foller}},\ and\ \bibinfo {author}
		{\bibfnamefont {M.}~\bibnamefont {Schleberger}},\ }\href
	{https://doi.org/10.1088/2053-1583/aa91a7} {\bibfield  {journal} {\bibinfo
			{journal} {2D Mater.}\ }\textbf {\bibinfo {volume} {5}},\ \bibinfo {pages}
		{015014} (\bibinfo {year} {2017})}\BibitemShut {NoStop}%
	\bibitem [{\citenamefont {Shi}\ \emph {et~al.}(2016)\citenamefont {Shi},
		\citenamefont {Tong}, \citenamefont {Zhou}, \citenamefont {Gong},
		\citenamefont {Zhang}, \citenamefont {Ji}, \citenamefont {Zhang},
		\citenamefont {Fang}, \citenamefont {Gu}, \citenamefont {Wang}, \citenamefont
		{Liu},\ and\ \citenamefont {Zhang}}]{shi_temperature-mediated_2016}%
	\BibitemOpen
	\bibfield  {author} {\bibinfo {author} {\bibfnamefont {J.}~\bibnamefont
			{Shi}}, \bibinfo {author} {\bibfnamefont {R.}~\bibnamefont {Tong}}, \bibinfo
		{author} {\bibfnamefont {X.}~\bibnamefont {Zhou}}, \bibinfo {author}
		{\bibfnamefont {Y.}~\bibnamefont {Gong}}, \bibinfo {author} {\bibfnamefont
			{Z.}~\bibnamefont {Zhang}}, \bibinfo {author} {\bibfnamefont
			{Q.}~\bibnamefont {Ji}}, \bibinfo {author} {\bibfnamefont {Y.}~\bibnamefont
			{Zhang}}, \bibinfo {author} {\bibfnamefont {Q.}~\bibnamefont {Fang}},
		\bibinfo {author} {\bibfnamefont {L.}~\bibnamefont {Gu}}, \bibinfo {author}
		{\bibfnamefont {X.}~\bibnamefont {Wang}}, \bibinfo {author} {\bibfnamefont
			{Z.}~\bibnamefont {Liu}},\ and\ \bibinfo {author} {\bibfnamefont
			{Y.}~\bibnamefont {Zhang}},\ }\href {https://doi.org/10.1002/adma.201603174}
	{\bibfield  {journal} {\bibinfo  {journal} {Adv. Mater.}\ }\textbf {\bibinfo
			{volume} {28}},\ \bibinfo {pages} {10664} (\bibinfo {year}
		{2016})}\BibitemShut {NoStop}%
	\bibitem [{\citenamefont {Guo}\ \emph {et~al.}(2020)\citenamefont {Guo},
		\citenamefont {Li}, \citenamefont {Liu}, \citenamefont {Jin}, \citenamefont
		{Chen}, \citenamefont {Guo},\ and\ \citenamefont {Lian}}]{guo_enhanced_2020}%
	\BibitemOpen
	\bibfield  {author} {\bibinfo {author} {\bibfnamefont {X.}~\bibnamefont
			{Guo}}, \bibinfo {author} {\bibfnamefont {Q.}~\bibnamefont {Li}}, \bibinfo
		{author} {\bibfnamefont {Y.}~\bibnamefont {Liu}}, \bibinfo {author}
		{\bibfnamefont {T.}~\bibnamefont {Jin}}, \bibinfo {author} {\bibfnamefont
			{Y.}~\bibnamefont {Chen}}, \bibinfo {author} {\bibfnamefont {L.}~\bibnamefont
			{Guo}},\ and\ \bibinfo {author} {\bibfnamefont {T.}~\bibnamefont {Lian}},\
	}\href {https://doi.org/10.1021/acsami.0c12931} {\bibfield  {journal}
		{\bibinfo  {journal} {ACS Appl. Mater. Interfaces}\ }\textbf {\bibinfo
			{volume} {12}},\ \bibinfo {pages} {44769} (\bibinfo {year}
		{2020})}\BibitemShut {NoStop}%
	\bibitem [{\citenamefont {Strange}\ \emph {et~al.}(2020)\citenamefont
		{Strange}, \citenamefont {Yadav}, \citenamefont {Garg}, \citenamefont
		{Shinde}, \citenamefont {Hill}, \citenamefont {Hill}, \citenamefont {Kung},\
		and\ \citenamefont {Pan}}]{strange_investigating_2020}%
	\BibitemOpen
	\bibfield  {author} {\bibinfo {author} {\bibfnamefont {L.~E.}\ \bibnamefont
			{Strange}}, \bibinfo {author} {\bibfnamefont {J.}~\bibnamefont {Yadav}},
		\bibinfo {author} {\bibfnamefont {S.}~\bibnamefont {Garg}}, \bibinfo {author}
		{\bibfnamefont {P.~S.}\ \bibnamefont {Shinde}}, \bibinfo {author}
		{\bibfnamefont {J.~W.}\ \bibnamefont {Hill}}, \bibinfo {author}
		{\bibfnamefont {C.~M.}\ \bibnamefont {Hill}}, \bibinfo {author}
		{\bibfnamefont {P.}~\bibnamefont {Kung}},\ and\ \bibinfo {author}
		{\bibfnamefont {S.}~\bibnamefont {Pan}},\ }\href
	{https://doi.org/10.1021/acs.jpclett.0c00769} {\bibfield  {journal} {\bibinfo
			{journal} {J. Phys. Chem. Lett.}\ }\textbf {\bibinfo {volume} {11}},\
		\bibinfo {pages} {3488} (\bibinfo {year} {2020})}\BibitemShut {NoStop}%
	\bibitem [{\citenamefont {Bruix}\ \emph {et~al.}(2016)\citenamefont {Bruix},
		\citenamefont {Miwa}, \citenamefont {Hauptmann}, \citenamefont {Wegner},
		\citenamefont {Ulstrup}, \citenamefont {Grønborg}, \citenamefont {Sanders},
		\citenamefont {Dendzik}, \citenamefont {Grubišić~Čabo}, \citenamefont
		{Bianchi}, \citenamefont {Lauritsen}, \citenamefont {Khajetoorians},
		\citenamefont {Hammer},\ and\ \citenamefont
		{Hofmann}}]{bruix_single-layer_2016}%
	\BibitemOpen
	\bibfield  {author} {\bibinfo {author} {\bibfnamefont {A.}~\bibnamefont
			{Bruix}}, \bibinfo {author} {\bibfnamefont {J.~A.}\ \bibnamefont {Miwa}},
		\bibinfo {author} {\bibfnamefont {N.}~\bibnamefont {Hauptmann}}, \bibinfo
		{author} {\bibfnamefont {D.}~\bibnamefont {Wegner}}, \bibinfo {author}
		{\bibfnamefont {S.}~\bibnamefont {Ulstrup}}, \bibinfo {author} {\bibfnamefont
			{S.~S.}\ \bibnamefont {Grønborg}}, \bibinfo {author} {\bibfnamefont {C.~E.}\
			\bibnamefont {Sanders}}, \bibinfo {author} {\bibfnamefont {M.}~\bibnamefont
			{Dendzik}}, \bibinfo {author} {\bibfnamefont {A.}~\bibnamefont
			{Grubišić~Čabo}}, \bibinfo {author} {\bibfnamefont {M.}~\bibnamefont
			{Bianchi}}, \bibinfo {author} {\bibfnamefont {J.~V.}\ \bibnamefont
			{Lauritsen}}, \bibinfo {author} {\bibfnamefont {A.~A.}\ \bibnamefont
			{Khajetoorians}}, \bibinfo {author} {\bibfnamefont {B.}~\bibnamefont
			{Hammer}},\ and\ \bibinfo {author} {\bibfnamefont {P.}~\bibnamefont
			{Hofmann}},\ }\href {https://doi.org/10.1103/PhysRevB.93.165422} {\bibfield
		{journal} {\bibinfo  {journal} {Phys. Rev. B}\ }\textbf {\bibinfo {volume}
			{93}},\ \bibinfo {pages} {165422} (\bibinfo {year} {2016})}\BibitemShut
	{NoStop}%
	\bibitem [{\citenamefont {Silva}\ \emph {et~al.}(2022)\citenamefont {Silva},
		\citenamefont {Dombrowski}, \citenamefont {Atodiresei}, \citenamefont
		{Jolie}, \citenamefont {Hagen}, \citenamefont {Cai}, \citenamefont {Ryan},
		\citenamefont {Thakur}, \citenamefont {Caciuc}, \citenamefont {Blügel},
		\citenamefont {Duncan}, \citenamefont {Michely}, \citenamefont {Lee},\ and\
		\citenamefont {Busse}}]{silva_spatial_2022}%
	\BibitemOpen
	\bibfield  {author} {\bibinfo {author} {\bibfnamefont {C.~C.}\ \bibnamefont
			{Silva}}, \bibinfo {author} {\bibfnamefont {D.}~\bibnamefont {Dombrowski}},
		\bibinfo {author} {\bibfnamefont {N.}~\bibnamefont {Atodiresei}}, \bibinfo
		{author} {\bibfnamefont {W.}~\bibnamefont {Jolie}}, \bibinfo {author}
		{\bibfnamefont {F.~F.~z.}\ \bibnamefont {Hagen}}, \bibinfo {author}
		{\bibfnamefont {J.}~\bibnamefont {Cai}}, \bibinfo {author} {\bibfnamefont
			{P.~T.~P.}\ \bibnamefont {Ryan}}, \bibinfo {author} {\bibfnamefont {P.~K.}\
			\bibnamefont {Thakur}}, \bibinfo {author} {\bibfnamefont {V.}~\bibnamefont
			{Caciuc}}, \bibinfo {author} {\bibfnamefont {S.}~\bibnamefont {Blügel}},
		\bibinfo {author} {\bibfnamefont {D.~A.}\ \bibnamefont {Duncan}}, \bibinfo
		{author} {\bibfnamefont {T.}~\bibnamefont {Michely}}, \bibinfo {author}
		{\bibfnamefont {T.-L.}\ \bibnamefont {Lee}},\ and\ \bibinfo {author}
		{\bibfnamefont {C.}~\bibnamefont {Busse}},\ }\href
	{https://doi.org/10.1088/2053-1583/ac4958} {\bibfield  {journal} {\bibinfo
			{journal} {2D Mater.}\ }\textbf {\bibinfo {volume} {9}},\ \bibinfo {pages}
		{025003} (\bibinfo {year} {2022})}\BibitemShut {NoStop}%
	\bibitem [{\citenamefont {Wang}\ \emph {et~al.}(2018)\citenamefont {Wang},
		\citenamefont {Chernikov}, \citenamefont {Glazov}, \citenamefont {Heinz},
		\citenamefont {Marie}, \citenamefont {Amand},\ and\ \citenamefont
		{Urbaszek}}]{wang_colloquium_2018}%
	\BibitemOpen
	\bibfield  {author} {\bibinfo {author} {\bibfnamefont {G.}~\bibnamefont
			{Wang}}, \bibinfo {author} {\bibfnamefont {A.}~\bibnamefont {Chernikov}},
		\bibinfo {author} {\bibfnamefont {M.~M.}\ \bibnamefont {Glazov}}, \bibinfo
		{author} {\bibfnamefont {T.~F.}\ \bibnamefont {Heinz}}, \bibinfo {author}
		{\bibfnamefont {X.}~\bibnamefont {Marie}}, \bibinfo {author} {\bibfnamefont
			{T.}~\bibnamefont {Amand}},\ and\ \bibinfo {author} {\bibfnamefont
			{B.}~\bibnamefont {Urbaszek}},\ }\href
	{https://doi.org/10.1103/RevModPhys.90.021001} {\bibfield  {journal}
		{\bibinfo  {journal} {Rev. Mod. Phys.}\ }\textbf {\bibinfo {volume} {90}},\
		\bibinfo {pages} {021001} (\bibinfo {year} {2018})}\BibitemShut {NoStop}%
	\bibitem [{\citenamefont {Čabo}\ \emph {et~al.}(2015)\citenamefont {Čabo},
		\citenamefont {Miwa}, \citenamefont {Grønborg}, \citenamefont {Riley},
		\citenamefont {Johannsen}, \citenamefont {Cacho}, \citenamefont {Alexander},
		\citenamefont {Chapman}, \citenamefont {Springate}, \citenamefont {Grioni},
		\citenamefont {Lauritsen}, \citenamefont {King}, \citenamefont {Hofmann},\
		and\ \citenamefont {Ulstrup}}]{cabo_observation_2015}%
	\BibitemOpen
	\bibfield  {author} {\bibinfo {author} {\bibfnamefont {A.~G.}\ \bibnamefont
			{Čabo}}, \bibinfo {author} {\bibfnamefont {J.~A.}\ \bibnamefont {Miwa}},
		\bibinfo {author} {\bibfnamefont {S.~S.}\ \bibnamefont {Grønborg}}, \bibinfo
		{author} {\bibfnamefont {J.~M.}\ \bibnamefont {Riley}}, \bibinfo {author}
		{\bibfnamefont {J.~C.}\ \bibnamefont {Johannsen}}, \bibinfo {author}
		{\bibfnamefont {C.}~\bibnamefont {Cacho}}, \bibinfo {author} {\bibfnamefont
			{O.}~\bibnamefont {Alexander}}, \bibinfo {author} {\bibfnamefont {R.~T.}\
			\bibnamefont {Chapman}}, \bibinfo {author} {\bibfnamefont {E.}~\bibnamefont
			{Springate}}, \bibinfo {author} {\bibfnamefont {M.}~\bibnamefont {Grioni}},
		\bibinfo {author} {\bibfnamefont {J.~V.}\ \bibnamefont {Lauritsen}}, \bibinfo
		{author} {\bibfnamefont {P.~D.~C.}\ \bibnamefont {King}}, \bibinfo {author}
		{\bibfnamefont {P.}~\bibnamefont {Hofmann}},\ and\ \bibinfo {author}
		{\bibfnamefont {S.}~\bibnamefont {Ulstrup}},\ }\href
	{https://doi.org/10.1021/acs.nanolett.5b01967} {\bibfield  {journal}
		{\bibinfo  {journal} {Nano Lett.}\ }\textbf {\bibinfo {volume} {15}},\
		\bibinfo {pages} {5883} (\bibinfo {year} {2015})}\BibitemShut {NoStop}%
	\bibitem [{\citenamefont {Liu}\ \emph {et~al.}(2019)\citenamefont {Liu},
		\citenamefont {Ziffer}, \citenamefont {Hansen}, \citenamefont {Wang},\ and\
		\citenamefont {Zhu}}]{liu_direct_2019}%
	\BibitemOpen
	\bibfield  {author} {\bibinfo {author} {\bibfnamefont {F.}~\bibnamefont
			{Liu}}, \bibinfo {author} {\bibfnamefont {M.~E.}\ \bibnamefont {Ziffer}},
		\bibinfo {author} {\bibfnamefont {K.~R.}\ \bibnamefont {Hansen}}, \bibinfo
		{author} {\bibfnamefont {J.}~\bibnamefont {Wang}},\ and\ \bibinfo {author}
		{\bibfnamefont {X.}~\bibnamefont {Zhu}},\ }\href
	{https://doi.org/10.1103/PhysRevLett.122.246803} {\bibfield  {journal}
		{\bibinfo  {journal} {Phys. Rev. Lett.}\ }\textbf {\bibinfo {volume} {122}},\
		\bibinfo {pages} {246803} (\bibinfo {year} {2019})}\BibitemShut {NoStop}%
	\bibitem [{\citenamefont {Tauc}(1968)}]{tauc_optical_1968}%
	\BibitemOpen
	\bibfield  {author} {\bibinfo {author} {\bibfnamefont {J.}~\bibnamefont
			{Tauc}},\ }\href {https://doi.org/10.1016/0025-5408(68)90023-8} {\bibfield
		{journal} {\bibinfo  {journal} {Materials Research Bulletin}\ }\textbf
		{\bibinfo {volume} {3}},\ \bibinfo {pages} {37} (\bibinfo {year}
		{1968})}\BibitemShut {NoStop}%
	\bibitem [{\citenamefont {Klein}\ \emph {et~al.}(2023)\citenamefont {Klein},
		\citenamefont {Kampermann}, \citenamefont {Mockenhaupt}, \citenamefont
		{Behrens}, \citenamefont {Strunk},\ and\ \citenamefont
		{Bacher}}]{klein_limitations_2023}%
	\BibitemOpen
	\bibfield  {author} {\bibinfo {author} {\bibfnamefont {J.}~\bibnamefont
			{Klein}}, \bibinfo {author} {\bibfnamefont {L.}~\bibnamefont {Kampermann}},
		\bibinfo {author} {\bibfnamefont {B.}~\bibnamefont {Mockenhaupt}}, \bibinfo
		{author} {\bibfnamefont {M.}~\bibnamefont {Behrens}}, \bibinfo {author}
		{\bibfnamefont {J.}~\bibnamefont {Strunk}},\ and\ \bibinfo {author}
		{\bibfnamefont {G.}~\bibnamefont {Bacher}},\ }\href
	{https://doi.org/10.1002/adfm.202304523} {\bibfield  {journal} {\bibinfo
			{journal} {Adv. Funct. Mater.}\ }\textbf {\bibinfo {volume} {n/a}},\ \bibinfo
		{pages} {2304523} (\bibinfo {year} {2023})}\BibitemShut {NoStop}%
	\bibitem [{\citenamefont {Grundmann}(2016)}]{grundmann_physics_2016}%
	\BibitemOpen
	\bibfield  {author} {\bibinfo {author} {\bibfnamefont {M.}~\bibnamefont
			{Grundmann}},\ }\href {https://link.springer.com/10.1007/978-3-319-23880-7}
	{\emph {\bibinfo {title} {The {Physics} of {Semiconductors}: {An}
				{Introduction} {Including} {Nanophysics} and {Applications}}}}\ (\bibinfo
	{publisher} {Springer International Publishing},\ \bibinfo {year}
	{2016})\BibitemShut {NoStop}%
	\bibitem [{\citenamefont {Zheng}\ \emph {et~al.}(2020)\citenamefont {Zheng},
		\citenamefont {Bonn},\ and\ \citenamefont
		{Wang}}]{zheng_photoconductivity_2020}%
	\BibitemOpen
	\bibfield  {author} {\bibinfo {author} {\bibfnamefont {W.}~\bibnamefont
			{Zheng}}, \bibinfo {author} {\bibfnamefont {M.}~\bibnamefont {Bonn}},\ and\
		\bibinfo {author} {\bibfnamefont {H.~I.}\ \bibnamefont {Wang}},\ }\href
	{https://doi.org/10.1021/acs.nanolett.0c01693} {\bibfield  {journal}
		{\bibinfo  {journal} {Nano Lett.}\ }\textbf {\bibinfo {volume} {20}},\
		\bibinfo {pages} {5807} (\bibinfo {year} {2020})}\BibitemShut {NoStop}%
	\bibitem [{\citenamefont {Paul}\ and\ \citenamefont
		{Grinberg}(2022)}]{paul_optical_2022}%
	\BibitemOpen
	\bibfield  {author} {\bibinfo {author} {\bibfnamefont {A.}~\bibnamefont
			{Paul}}\ and\ \bibinfo {author} {\bibfnamefont {I.}~\bibnamefont
			{Grinberg}},\ }\href {https://doi.org/10.1103/PhysRevApplied.17.024042}
	{\bibfield  {journal} {\bibinfo  {journal} {Phys. Rev. Appl.}\ }\textbf
		{\bibinfo {volume} {17}},\ \bibinfo {pages} {024042} (\bibinfo {year}
		{2022})}\BibitemShut {NoStop}%
	\bibitem [{\citenamefont {Mattinen}\ \emph {et~al.}(2017)\citenamefont
		{Mattinen}, \citenamefont {Hatanpää}, \citenamefont {Sarnet}, \citenamefont
		{Mizohata}, \citenamefont {Meinander}, \citenamefont {King}, \citenamefont
		{Khriachtchev}, \citenamefont {Räisänen}, \citenamefont {Ritala},\ and\
		\citenamefont {Leskelä}}]{mattinen_atomic_2017}%
	\BibitemOpen
	\bibfield  {author} {\bibinfo {author} {\bibfnamefont {M.}~\bibnamefont
			{Mattinen}}, \bibinfo {author} {\bibfnamefont {T.}~\bibnamefont
			{Hatanpää}}, \bibinfo {author} {\bibfnamefont {T.}~\bibnamefont {Sarnet}},
		\bibinfo {author} {\bibfnamefont {K.}~\bibnamefont {Mizohata}}, \bibinfo
		{author} {\bibfnamefont {K.}~\bibnamefont {Meinander}}, \bibinfo {author}
		{\bibfnamefont {P.~J.}\ \bibnamefont {King}}, \bibinfo {author}
		{\bibfnamefont {L.}~\bibnamefont {Khriachtchev}}, \bibinfo {author}
		{\bibfnamefont {J.}~\bibnamefont {Räisänen}}, \bibinfo {author}
		{\bibfnamefont {M.}~\bibnamefont {Ritala}},\ and\ \bibinfo {author}
		{\bibfnamefont {M.}~\bibnamefont {Leskelä}},\ }\href
	{https://doi.org/10.1002/admi.201700123} {\bibfield  {journal} {\bibinfo
			{journal} {Adv. Mater. Interfaces}\ }\textbf {\bibinfo {volume} {4}},\
		\bibinfo {pages} {1700123} (\bibinfo {year} {2017})}\BibitemShut {NoStop}%
	\bibitem [{\citenamefont {Wang}\ \emph {et~al.}(2021)\citenamefont {Wang},
		\citenamefont {Huang}, \citenamefont {Li}, \citenamefont {Azam},
		\citenamefont {Zu}, \citenamefont {Qiao}, \citenamefont {Yang},\ and\
		\citenamefont {Li}}]{wang_growth_2021}%
	\BibitemOpen
	\bibfield  {author} {\bibinfo {author} {\bibfnamefont {S.}~\bibnamefont
			{Wang}}, \bibinfo {author} {\bibfnamefont {J.-K.}\ \bibnamefont {Huang}},
		\bibinfo {author} {\bibfnamefont {M.}~\bibnamefont {Li}}, \bibinfo {author}
		{\bibfnamefont {A.}~\bibnamefont {Azam}}, \bibinfo {author} {\bibfnamefont
			{X.}~\bibnamefont {Zu}}, \bibinfo {author} {\bibfnamefont {L.}~\bibnamefont
			{Qiao}}, \bibinfo {author} {\bibfnamefont {J.}~\bibnamefont {Yang}},\ and\
		\bibinfo {author} {\bibfnamefont {S.}~\bibnamefont {Li}},\ }\href
	{https://doi.org/10.1021/acsami.1c14136} {\bibfield  {journal} {\bibinfo
			{journal} {ACS Appl. Mater. Interfaces}\ }\textbf {\bibinfo {volume} {13}},\
		\bibinfo {pages} {47962} (\bibinfo {year} {2021})}\BibitemShut {NoStop}%
	\bibitem [{\citenamefont {Bhanu}\ \emph {et~al.}(2015)\citenamefont {Bhanu},
		\citenamefont {Islam}, \citenamefont {Tetard},\ and\ \citenamefont
		{Khondaker}}]{bhanu_photoluminescence_2015}%
	\BibitemOpen
	\bibfield  {author} {\bibinfo {author} {\bibfnamefont {U.}~\bibnamefont
			{Bhanu}}, \bibinfo {author} {\bibfnamefont {M.~R.}\ \bibnamefont {Islam}},
		\bibinfo {author} {\bibfnamefont {L.}~\bibnamefont {Tetard}},\ and\ \bibinfo
		{author} {\bibfnamefont {S.~I.}\ \bibnamefont {Khondaker}},\ }\href
	{https://doi.org/10.1038/srep05575} {\bibfield  {journal} {\bibinfo
			{journal} {Sci. Rep.}\ }\textbf {\bibinfo {volume} {4}},\ \bibinfo {pages}
		{5575} (\bibinfo {year} {2015})}\BibitemShut {NoStop}%
	\bibitem [{\citenamefont {Pollmann}\ \emph {et~al.}(2021)\citenamefont
		{Pollmann}, \citenamefont {Sleziona}, \citenamefont {Foller}, \citenamefont
		{Hagemann}, \citenamefont {Gorynski}, \citenamefont {Petri}, \citenamefont
		{Madauß}, \citenamefont {Breuer},\ and\ \citenamefont
		{Schleberger}}]{pollmann_large-area_2021}%
	\BibitemOpen
	\bibfield  {author} {\bibinfo {author} {\bibfnamefont {E.}~\bibnamefont
			{Pollmann}}, \bibinfo {author} {\bibfnamefont {S.}~\bibnamefont {Sleziona}},
		\bibinfo {author} {\bibfnamefont {T.}~\bibnamefont {Foller}}, \bibinfo
		{author} {\bibfnamefont {U.}~\bibnamefont {Hagemann}}, \bibinfo {author}
		{\bibfnamefont {C.}~\bibnamefont {Gorynski}}, \bibinfo {author}
		{\bibfnamefont {O.}~\bibnamefont {Petri}}, \bibinfo {author} {\bibfnamefont
			{L.}~\bibnamefont {Madauß}}, \bibinfo {author} {\bibfnamefont
			{L.}~\bibnamefont {Breuer}},\ and\ \bibinfo {author} {\bibfnamefont
			{M.}~\bibnamefont {Schleberger}},\ }\href
	{https://doi.org/10.1021/acsomega.1c01570} {\bibfield  {journal} {\bibinfo
			{journal} {ACS Omega}\ }\textbf {\bibinfo {volume} {6}},\ \bibinfo {pages}
		{15929} (\bibinfo {year} {2021})}\BibitemShut {NoStop}%
	\bibitem [{\citenamefont {Zou}\ \emph {et~al.}(2021)\citenamefont {Zou},
		\citenamefont {Wu}, \citenamefont {Zhou}, \citenamefont {Zhou}, \citenamefont
		{Wang}, \citenamefont {Zhang}, \citenamefont {Cao},\ and\ \citenamefont
		{Sun}}]{zou_spectroscopic_2021}%
	\BibitemOpen
	\bibfield  {author} {\bibinfo {author} {\bibfnamefont {B.}~\bibnamefont
			{Zou}}, \bibinfo {author} {\bibfnamefont {Z.}~\bibnamefont {Wu}}, \bibinfo
		{author} {\bibfnamefont {Y.}~\bibnamefont {Zhou}}, \bibinfo {author}
		{\bibfnamefont {Y.}~\bibnamefont {Zhou}}, \bibinfo {author} {\bibfnamefont
			{J.}~\bibnamefont {Wang}}, \bibinfo {author} {\bibfnamefont {L.}~\bibnamefont
			{Zhang}}, \bibinfo {author} {\bibfnamefont {F.}~\bibnamefont {Cao}},\ and\
		\bibinfo {author} {\bibfnamefont {H.}~\bibnamefont {Sun}},\ }\href
	{https://doi.org/10.1002/pssr.202100385} {\bibfield  {journal} {\bibinfo
			{journal} {Phys. Status Solidi RRL}\ }\textbf {\bibinfo {volume} {15}},\
		\bibinfo {pages} {2100385} (\bibinfo {year} {2021})}\BibitemShut {NoStop}%
	\bibitem [{\citenamefont {Park}\ \emph {et~al.}(2018)\citenamefont {Park},
		\citenamefont {Mutz}, \citenamefont {Schultz}, \citenamefont {Blumstengel},
		\citenamefont {Han}, \citenamefont {Aljarb}, \citenamefont {Li},
		\citenamefont {List-Kratochvil}, \citenamefont {Amsalem},\ and\ \citenamefont
		{Koch}}]{park_direct_2018}%
	\BibitemOpen
	\bibfield  {author} {\bibinfo {author} {\bibfnamefont {S.}~\bibnamefont
			{Park}}, \bibinfo {author} {\bibfnamefont {N.}~\bibnamefont {Mutz}}, \bibinfo
		{author} {\bibfnamefont {T.}~\bibnamefont {Schultz}}, \bibinfo {author}
		{\bibfnamefont {S.}~\bibnamefont {Blumstengel}}, \bibinfo {author}
		{\bibfnamefont {A.}~\bibnamefont {Han}}, \bibinfo {author} {\bibfnamefont
			{A.}~\bibnamefont {Aljarb}}, \bibinfo {author} {\bibfnamefont {L.-J.}\
			\bibnamefont {Li}}, \bibinfo {author} {\bibfnamefont {E.~J.~W.}\ \bibnamefont
			{List-Kratochvil}}, \bibinfo {author} {\bibfnamefont {P.}~\bibnamefont
			{Amsalem}},\ and\ \bibinfo {author} {\bibfnamefont {N.}~\bibnamefont
			{Koch}},\ }\href {https://doi.org/10.1088/2053-1583/aaa4ca} {\bibfield
		{journal} {\bibinfo  {journal} {2D Mater.}\ }\textbf {\bibinfo {volume}
			{5}},\ \bibinfo {pages} {025003} (\bibinfo {year} {2018})}\BibitemShut
	{NoStop}%
	\bibitem [{\citenamefont {Shen}(1989)}]{shen_surface_1989}%
	\BibitemOpen
	\bibfield  {author} {\bibinfo {author} {\bibfnamefont {Y.~R.}\ \bibnamefont
			{Shen}},\ }\href {https://doi.org/10.1038/337519a0} {\bibfield  {journal}
		{\bibinfo  {journal} {Nature}\ }\textbf {\bibinfo {volume} {337}},\ \bibinfo
		{pages} {519} (\bibinfo {year} {1989})}\BibitemShut {NoStop}%
	\bibitem [{\citenamefont {Boyd}(2020)}]{boyd_nonlinear_2020}%
	\BibitemOpen
	\bibfield  {author} {\bibinfo {author} {\bibfnamefont {R.~W.}\ \bibnamefont
			{Boyd}},\ }\href@noop {} {\emph {\bibinfo {title} {Nonlinear {Optics}}}}\
	(\bibinfo  {publisher} {Academic Press},\ \bibinfo {year} {2020})\BibitemShut
	{NoStop}%
	\bibitem [{\citenamefont {Yang}\ \emph {et~al.}(2023)\citenamefont {Yang},
		\citenamefont {Pollmann}, \citenamefont {Sleziona}, \citenamefont
		{Hasselbrink}, \citenamefont {Kratzer}, \citenamefont {Schleberger},
		\citenamefont {Campen},\ and\ \citenamefont {Tong}}]{yang_interaction_2023}%
	\BibitemOpen
	\bibfield  {author} {\bibinfo {author} {\bibfnamefont {T.}~\bibnamefont
			{Yang}}, \bibinfo {author} {\bibfnamefont {E.}~\bibnamefont {Pollmann}},
		\bibinfo {author} {\bibfnamefont {S.}~\bibnamefont {Sleziona}}, \bibinfo
		{author} {\bibfnamefont {E.}~\bibnamefont {Hasselbrink}}, \bibinfo {author}
		{\bibfnamefont {P.}~\bibnamefont {Kratzer}}, \bibinfo {author} {\bibfnamefont
			{M.}~\bibnamefont {Schleberger}}, \bibinfo {author} {\bibfnamefont {R.~K.}\
			\bibnamefont {Campen}},\ and\ \bibinfo {author} {\bibfnamefont
			{Y.}~\bibnamefont {Tong}},\ }\href
	{https://doi.org/10.1103/PhysRevB.107.155433} {\bibfield  {journal} {\bibinfo
			{journal} {Phys. Rev. B}\ }\textbf {\bibinfo {volume} {107}},\ \bibinfo
		{pages} {155433} (\bibinfo {year} {2023})}\BibitemShut {NoStop}%
	\bibitem [{Yan()}]{Yang_spectrum_2023}%
	\BibitemOpen
	\href@noop {} {\ }\BibitemShut {NoStop}%
	\bibitem [{\citenamefont {Tong}\ \emph {et~al.}(2017)\citenamefont {Tong},
		\citenamefont {Lapointe}, \citenamefont {Thämer}, \citenamefont {Wolf},\
		and\ \citenamefont {Campen}}]{tong_hydrophobic_2017}%
	\BibitemOpen
	\bibfield  {author} {\bibinfo {author} {\bibfnamefont {Y.}~\bibnamefont
			{Tong}}, \bibinfo {author} {\bibfnamefont {F.}~\bibnamefont {Lapointe}},
		\bibinfo {author} {\bibfnamefont {M.}~\bibnamefont {Thämer}}, \bibinfo
		{author} {\bibfnamefont {M.}~\bibnamefont {Wolf}},\ and\ \bibinfo {author}
		{\bibfnamefont {R.~K.}\ \bibnamefont {Campen}},\ }\href
	{https://doi.org/10.1002/anie.201612183} {\bibfield  {journal} {\bibinfo
			{journal} {Angew. Chem., Int. Ed.}\ }\textbf {\bibinfo {volume} {56}},\
		\bibinfo {pages} {4211} (\bibinfo {year} {2017})}\BibitemShut {NoStop}%
	\bibitem [{\citenamefont {Tong}\ \emph {et~al.}(2018)\citenamefont {Tong},
		\citenamefont {Zhang},\ and\ \citenamefont
		{Campen}}]{tong_experimentally_2018}%
	\BibitemOpen
	\bibfield  {author} {\bibinfo {author} {\bibfnamefont {Y.}~\bibnamefont
			{Tong}}, \bibinfo {author} {\bibfnamefont {I.~Y.}\ \bibnamefont {Zhang}},\
		and\ \bibinfo {author} {\bibfnamefont {R.~K.}\ \bibnamefont {Campen}},\
	}\href {https://doi.org/10.1038/s41467-018-03598-x} {\bibfield  {journal}
		{\bibinfo  {journal} {Nat. Commun.}\ }\textbf {\bibinfo {volume} {9}},\
		\bibinfo {pages} {1313} (\bibinfo {year} {2018})}\BibitemShut {NoStop}%
	\bibitem [{\citenamefont {Lee}\ \emph {et~al.}(2010)\citenamefont {Lee},
		\citenamefont {Yan}, \citenamefont {Brus}, \citenamefont {Heinz},
		\citenamefont {Hone},\ and\ \citenamefont {Ryu}}]{lee_anomalous_2010}%
	\BibitemOpen
	\bibfield  {author} {\bibinfo {author} {\bibfnamefont {C.}~\bibnamefont
			{Lee}}, \bibinfo {author} {\bibfnamefont {H.}~\bibnamefont {Yan}}, \bibinfo
		{author} {\bibfnamefont {L.~E.}\ \bibnamefont {Brus}}, \bibinfo {author}
		{\bibfnamefont {T.~F.}\ \bibnamefont {Heinz}}, \bibinfo {author}
		{\bibfnamefont {J.}~\bibnamefont {Hone}},\ and\ \bibinfo {author}
		{\bibfnamefont {S.}~\bibnamefont {Ryu}},\ }\href
	{https://doi.org/10.1021/nn1003937} {\bibfield  {journal} {\bibinfo
			{journal} {ACS Nano}\ }\textbf {\bibinfo {volume} {4}},\ \bibinfo {pages}
		{2695} (\bibinfo {year} {2010})}\BibitemShut {NoStop}%
	\bibitem [{\citenamefont {Velický}\ \emph {et~al.}(2020)\citenamefont
		{Velický}, \citenamefont {Rodriguez}, \citenamefont {Bouša}, \citenamefont
		{Krayev}, \citenamefont {Vondráček}, \citenamefont {Honolka}, \citenamefont
		{Ahmadi}, \citenamefont {Donnelly}, \citenamefont {Huang}, \citenamefont
		{Abruña}, \citenamefont {Novoselov},\ and\ \citenamefont
		{Frank}}]{velicky_strain_2020}%
	\BibitemOpen
	\bibfield  {author} {\bibinfo {author} {\bibfnamefont {M.}~\bibnamefont
			{Velický}}, \bibinfo {author} {\bibfnamefont {A.}~\bibnamefont {Rodriguez}},
		\bibinfo {author} {\bibfnamefont {M.}~\bibnamefont {Bouša}}, \bibinfo
		{author} {\bibfnamefont {A.~V.}\ \bibnamefont {Krayev}}, \bibinfo {author}
		{\bibfnamefont {M.}~\bibnamefont {Vondráček}}, \bibinfo {author}
		{\bibfnamefont {J.}~\bibnamefont {Honolka}}, \bibinfo {author} {\bibfnamefont
			{M.}~\bibnamefont {Ahmadi}}, \bibinfo {author} {\bibfnamefont {G.~E.}\
			\bibnamefont {Donnelly}}, \bibinfo {author} {\bibfnamefont {F.}~\bibnamefont
			{Huang}}, \bibinfo {author} {\bibfnamefont {H.~D.}\ \bibnamefont {Abruña}},
		\bibinfo {author} {\bibfnamefont {K.~S.}\ \bibnamefont {Novoselov}},\ and\
		\bibinfo {author} {\bibfnamefont {O.}~\bibnamefont {Frank}},\ }\href
	{https://doi.org/10.1021/acs.jpclett.0c01287} {\bibfield  {journal} {\bibinfo
			{journal} {J. Phys. Chem. Lett.}\ }\textbf {\bibinfo {volume} {11}},\
		\bibinfo {pages} {6112} (\bibinfo {year} {2020})}\BibitemShut {NoStop}%
	\bibitem [{\citenamefont {Sarkar}\ and\ \citenamefont
		{Kratzer}(2021)}]{sarkar_signatures_2021}%
	\BibitemOpen
	\bibfield  {author} {\bibinfo {author} {\bibfnamefont {S.}~\bibnamefont
			{Sarkar}}\ and\ \bibinfo {author} {\bibfnamefont {P.}~\bibnamefont
			{Kratzer}},\ }\href {https://doi.org/10.1021/acs.jpcc.1c08594} {\bibfield
		{journal} {\bibinfo  {journal} {J. Phys. Chem. C}\ }\textbf {\bibinfo
			{volume} {125}},\ \bibinfo {pages} {26645} (\bibinfo {year}
		{2021})}\BibitemShut {NoStop}%
	\bibitem [{\citenamefont {Xiao}\ \emph {et~al.}(2012)\citenamefont {Xiao},
		\citenamefont {Liu}, \citenamefont {Feng}, \citenamefont {Xu},\ and\
		\citenamefont {Yao}}]{xiao_coupled_2012}%
	\BibitemOpen
	\bibfield  {author} {\bibinfo {author} {\bibfnamefont {D.}~\bibnamefont
			{Xiao}}, \bibinfo {author} {\bibfnamefont {G.-B.}\ \bibnamefont {Liu}},
		\bibinfo {author} {\bibfnamefont {W.}~\bibnamefont {Feng}}, \bibinfo {author}
		{\bibfnamefont {X.}~\bibnamefont {Xu}},\ and\ \bibinfo {author}
		{\bibfnamefont {W.}~\bibnamefont {Yao}},\ }\href
	{https://doi.org/10.1103/PhysRevLett.108.196802} {\bibfield  {journal}
		{\bibinfo  {journal} {Phys. Rev. Lett.}\ }\textbf {\bibinfo {volume} {108}},\
		\bibinfo {pages} {196802} (\bibinfo {year} {2012})}\BibitemShut {NoStop}%
	\bibitem [{\citenamefont {Scheuschner}\ \emph {et~al.}(2014)\citenamefont
		{Scheuschner}, \citenamefont {Ochedowski}, \citenamefont {Kaulitz},
		\citenamefont {Gillen}, \citenamefont {Schleberger},\ and\ \citenamefont
		{Maultzsch}}]{scheuschner_photoluminescence_2014}%
	\BibitemOpen
	\bibfield  {author} {\bibinfo {author} {\bibfnamefont {N.}~\bibnamefont
			{Scheuschner}}, \bibinfo {author} {\bibfnamefont {O.}~\bibnamefont
			{Ochedowski}}, \bibinfo {author} {\bibfnamefont {A.-M.}\ \bibnamefont
			{Kaulitz}}, \bibinfo {author} {\bibfnamefont {R.}~\bibnamefont {Gillen}},
		\bibinfo {author} {\bibfnamefont {M.}~\bibnamefont {Schleberger}},\ and\
		\bibinfo {author} {\bibfnamefont {J.}~\bibnamefont {Maultzsch}},\ }\href
	{https://doi.org/10.1103/PhysRevB.89.125406} {\bibfield  {journal} {\bibinfo
			{journal} {Phys. Rev. B}\ }\textbf {\bibinfo {volume} {89}},\ \bibinfo
		{pages} {125406} (\bibinfo {year} {2014})}\BibitemShut {NoStop}%
	\bibitem [{\citenamefont {Kumar}\ \emph {et~al.}(2013)\citenamefont {Kumar},
		\citenamefont {Najmaei}, \citenamefont {Cui}, \citenamefont {Ceballos},
		\citenamefont {Ajayan}, \citenamefont {Lou},\ and\ \citenamefont
		{Zhao}}]{kumar_second_2013}%
	\BibitemOpen
	\bibfield  {author} {\bibinfo {author} {\bibfnamefont {N.}~\bibnamefont
			{Kumar}}, \bibinfo {author} {\bibfnamefont {S.}~\bibnamefont {Najmaei}},
		\bibinfo {author} {\bibfnamefont {Q.}~\bibnamefont {Cui}}, \bibinfo {author}
		{\bibfnamefont {F.}~\bibnamefont {Ceballos}}, \bibinfo {author}
		{\bibfnamefont {P.~M.}\ \bibnamefont {Ajayan}}, \bibinfo {author}
		{\bibfnamefont {J.}~\bibnamefont {Lou}},\ and\ \bibinfo {author}
		{\bibfnamefont {H.}~\bibnamefont {Zhao}},\ }\href
	{https://doi.org/10.1103/PhysRevB.87.161403} {\bibfield  {journal} {\bibinfo
			{journal} {Phys. Rev. B}\ }\textbf {\bibinfo {volume} {87}},\ \bibinfo
		{pages} {161403} (\bibinfo {year} {2013})}\BibitemShut {NoStop}%
	\bibitem [{\citenamefont {Malard}\ \emph {et~al.}(2013)\citenamefont {Malard},
		\citenamefont {Alencar}, \citenamefont {Barboza}, \citenamefont {Mak},\ and\
		\citenamefont {de~Paula}}]{malard_observation_2013}%
	\BibitemOpen
	\bibfield  {author} {\bibinfo {author} {\bibfnamefont {L.~M.}\ \bibnamefont
			{Malard}}, \bibinfo {author} {\bibfnamefont {T.~V.}\ \bibnamefont {Alencar}},
		\bibinfo {author} {\bibfnamefont {A.~P.~M.}\ \bibnamefont {Barboza}},
		\bibinfo {author} {\bibfnamefont {K.~F.}\ \bibnamefont {Mak}},\ and\ \bibinfo
		{author} {\bibfnamefont {A.~M.}\ \bibnamefont {de~Paula}},\ }\href
	{https://doi.org/10.1103/PhysRevB.87.201401} {\bibfield  {journal} {\bibinfo
			{journal} {Phys. Rev. B}\ }\textbf {\bibinfo {volume} {87}},\ \bibinfo
		{pages} {201401} (\bibinfo {year} {2013})}\BibitemShut {NoStop}%
	\bibitem [{\citenamefont {Jiang}\ \emph {et~al.}(2014)\citenamefont {Jiang},
		\citenamefont {Liu}, \citenamefont {Huang}, \citenamefont {Zhang},
		\citenamefont {Li}, \citenamefont {Gong}, \citenamefont {Shen}, \citenamefont
		{Liu},\ and\ \citenamefont {Wu}}]{jiang_valley_2014}%
	\BibitemOpen
	\bibfield  {author} {\bibinfo {author} {\bibfnamefont {T.}~\bibnamefont
			{Jiang}}, \bibinfo {author} {\bibfnamefont {H.}~\bibnamefont {Liu}}, \bibinfo
		{author} {\bibfnamefont {D.}~\bibnamefont {Huang}}, \bibinfo {author}
		{\bibfnamefont {S.}~\bibnamefont {Zhang}}, \bibinfo {author} {\bibfnamefont
			{Y.}~\bibnamefont {Li}}, \bibinfo {author} {\bibfnamefont {X.}~\bibnamefont
			{Gong}}, \bibinfo {author} {\bibfnamefont {Y.-R.}\ \bibnamefont {Shen}},
		\bibinfo {author} {\bibfnamefont {W.-T.}\ \bibnamefont {Liu}},\ and\ \bibinfo
		{author} {\bibfnamefont {S.}~\bibnamefont {Wu}},\ }\href
	{https://doi.org/10.1038/nnano.2014.176} {\bibfield  {journal} {\bibinfo
			{journal} {Nat. Nanotechnol.}\ }\textbf {\bibinfo {volume} {9}},\ \bibinfo
		{pages} {825} (\bibinfo {year} {2014})}\BibitemShut {NoStop}%
	\bibitem [{\citenamefont {Olsen}\ \emph {et~al.}(2016)\citenamefont {Olsen},
		\citenamefont {Latini}, \citenamefont {Rasmussen},\ and\ \citenamefont
		{Thygesen}}]{olsen_simple_2016}%
	\BibitemOpen
	\bibfield  {author} {\bibinfo {author} {\bibfnamefont {T.}~\bibnamefont
			{Olsen}}, \bibinfo {author} {\bibfnamefont {S.}~\bibnamefont {Latini}},
		\bibinfo {author} {\bibfnamefont {F.}~\bibnamefont {Rasmussen}},\ and\
		\bibinfo {author} {\bibfnamefont {K.~S.}\ \bibnamefont {Thygesen}},\ }\href
	{https://doi.org/10.1103/PhysRevLett.116.056401} {\bibfield  {journal}
		{\bibinfo  {journal} {Phys. Rev. Lett.}\ }\textbf {\bibinfo {volume} {116}},\
		\bibinfo {pages} {056401} (\bibinfo {year} {2016})}\BibitemShut {NoStop}%
	\bibitem [{\citenamefont {Drüppel}\ \emph {et~al.}(2017)\citenamefont
		{Drüppel}, \citenamefont {Deilmann}, \citenamefont {Krüger},\ and\
		\citenamefont {Rohlfing}}]{druppel_diversity_2017}%
	\BibitemOpen
	\bibfield  {author} {\bibinfo {author} {\bibfnamefont {M.}~\bibnamefont
			{Drüppel}}, \bibinfo {author} {\bibfnamefont {T.}~\bibnamefont {Deilmann}},
		\bibinfo {author} {\bibfnamefont {P.}~\bibnamefont {Krüger}},\ and\ \bibinfo
		{author} {\bibfnamefont {M.}~\bibnamefont {Rohlfing}},\ }\href
	{https://doi.org/10.1038/s41467-017-02286-6} {\bibfield  {journal} {\bibinfo
			{journal} {Nat. Commun.}\ }\textbf {\bibinfo {volume} {8}},\ \bibinfo {pages}
		{2117} (\bibinfo {year} {2017})}\BibitemShut {NoStop}%
	\bibitem [{\citenamefont {Steinhoff}\ \emph {et~al.}(2018)\citenamefont
		{Steinhoff}, \citenamefont {Wehling},\ and\ \citenamefont
		{Rösner}}]{steinhoff_frequency-dependent_2018}%
	\BibitemOpen
	\bibfield  {author} {\bibinfo {author} {\bibfnamefont {A.}~\bibnamefont
			{Steinhoff}}, \bibinfo {author} {\bibfnamefont {T.~O.}\ \bibnamefont
			{Wehling}},\ and\ \bibinfo {author} {\bibfnamefont {M.}~\bibnamefont
			{Rösner}},\ }\href {https://doi.org/10.1103/PhysRevB.98.045304} {\bibfield
		{journal} {\bibinfo  {journal} {Phys. Rev. B}\ }\textbf {\bibinfo {volume}
			{98}},\ \bibinfo {pages} {045304} (\bibinfo {year} {2018})}\BibitemShut
	{NoStop}%
	\bibitem [{\citenamefont {Ugeda}\ \emph {et~al.}(2014)\citenamefont {Ugeda},
		\citenamefont {Bradley}, \citenamefont {Shi}, \citenamefont {da~Jornada},
		\citenamefont {Zhang}, \citenamefont {Qiu}, \citenamefont {Ruan},
		\citenamefont {Mo}, \citenamefont {Hussain}, \citenamefont {Shen},
		\citenamefont {Wang}, \citenamefont {Louie},\ and\ \citenamefont
		{Crommie}}]{ugeda_giant_2014}%
	\BibitemOpen
	\bibfield  {author} {\bibinfo {author} {\bibfnamefont {M.~M.}\ \bibnamefont
			{Ugeda}}, \bibinfo {author} {\bibfnamefont {A.~J.}\ \bibnamefont {Bradley}},
		\bibinfo {author} {\bibfnamefont {S.-F.}\ \bibnamefont {Shi}}, \bibinfo
		{author} {\bibfnamefont {F.~H.}\ \bibnamefont {da~Jornada}}, \bibinfo
		{author} {\bibfnamefont {Y.}~\bibnamefont {Zhang}}, \bibinfo {author}
		{\bibfnamefont {D.~Y.}\ \bibnamefont {Qiu}}, \bibinfo {author} {\bibfnamefont
			{W.}~\bibnamefont {Ruan}}, \bibinfo {author} {\bibfnamefont {S.-K.}\
			\bibnamefont {Mo}}, \bibinfo {author} {\bibfnamefont {Z.}~\bibnamefont
			{Hussain}}, \bibinfo {author} {\bibfnamefont {Z.-X.}\ \bibnamefont {Shen}},
		\bibinfo {author} {\bibfnamefont {F.}~\bibnamefont {Wang}}, \bibinfo {author}
		{\bibfnamefont {S.~G.}\ \bibnamefont {Louie}},\ and\ \bibinfo {author}
		{\bibfnamefont {M.~F.}\ \bibnamefont {Crommie}},\ }\href
	{https://doi.org/10.1038/nmat4061} {\bibfield  {journal} {\bibinfo  {journal}
			{Nat. Mater.}\ }\textbf {\bibinfo {volume} {13}},\ \bibinfo {pages} {1091}
		(\bibinfo {year} {2014})}\BibitemShut {NoStop}%
	\bibitem [{\citenamefont {Cheiwchanchamnangij}\ and\ \citenamefont
		{Lambrecht}(2012)}]{cheiwchanchamnangij_quasiparticle_2012}%
	\BibitemOpen
	\bibfield  {author} {\bibinfo {author} {\bibfnamefont {T.}~\bibnamefont
			{Cheiwchanchamnangij}}\ and\ \bibinfo {author} {\bibfnamefont {W.~R.~L.}\
			\bibnamefont {Lambrecht}},\ }\href
	{https://doi.org/10.1103/PhysRevB.85.205302} {\bibfield  {journal} {\bibinfo
			{journal} {Phys. Rev. B}\ }\textbf {\bibinfo {volume} {85}},\ \bibinfo
		{pages} {205302} (\bibinfo {year} {2012})}\BibitemShut {NoStop}%
	\bibitem [{\citenamefont {Kormányos}\ \emph {et~al.}(2015)\citenamefont
		{Kormányos}, \citenamefont {Burkard}, \citenamefont {Gmitra}, \citenamefont
		{Fabian}, \citenamefont {Zólyomi}, \citenamefont {Drummond},\ and\
		\citenamefont {Fal’ko}}]{kormanyos_kp_2015}%
	\BibitemOpen
	\bibfield  {author} {\bibinfo {author} {\bibfnamefont {A.}~\bibnamefont
			{Kormányos}}, \bibinfo {author} {\bibfnamefont {G.}~\bibnamefont {Burkard}},
		\bibinfo {author} {\bibfnamefont {M.}~\bibnamefont {Gmitra}}, \bibinfo
		{author} {\bibfnamefont {J.}~\bibnamefont {Fabian}}, \bibinfo {author}
		{\bibfnamefont {V.}~\bibnamefont {Zólyomi}}, \bibinfo {author}
		{\bibfnamefont {N.~D.}\ \bibnamefont {Drummond}},\ and\ \bibinfo {author}
		{\bibfnamefont {V.}~\bibnamefont {Fal’ko}},\ }\href
	{https://doi.org/10.1088/2053-1583/2/2/022001} {\bibfield  {journal}
		{\bibinfo  {journal} {2D Mater.}\ }\textbf {\bibinfo {volume} {2}},\ \bibinfo
		{pages} {022001} (\bibinfo {year} {2015})}\BibitemShut {NoStop}%
	\bibitem [{\citenamefont {Wang}\ \emph {et~al.}(2019)\citenamefont {Wang},
		\citenamefont {Kim}, \citenamefont {Wu}, \citenamefont {Martinez},
		\citenamefont {Song}, \citenamefont {Yang}, \citenamefont {Zhao},
		\citenamefont {Mkhoyan}, \citenamefont {Jeong},\ and\ \citenamefont
		{Chhowalla}}]{wang_van_2019}%
	\BibitemOpen
	\bibfield  {author} {\bibinfo {author} {\bibfnamefont {Y.}~\bibnamefont
			{Wang}}, \bibinfo {author} {\bibfnamefont {J.~C.}\ \bibnamefont {Kim}},
		\bibinfo {author} {\bibfnamefont {R.~J.}\ \bibnamefont {Wu}}, \bibinfo
		{author} {\bibfnamefont {J.}~\bibnamefont {Martinez}}, \bibinfo {author}
		{\bibfnamefont {X.}~\bibnamefont {Song}}, \bibinfo {author} {\bibfnamefont
			{J.}~\bibnamefont {Yang}}, \bibinfo {author} {\bibfnamefont {F.}~\bibnamefont
			{Zhao}}, \bibinfo {author} {\bibfnamefont {A.}~\bibnamefont {Mkhoyan}},
		\bibinfo {author} {\bibfnamefont {H.~Y.}\ \bibnamefont {Jeong}},\ and\
		\bibinfo {author} {\bibfnamefont {M.}~\bibnamefont {Chhowalla}},\ }\href
	{https://doi.org/10.1038/s41586-019-1052-3} {\bibfield  {journal} {\bibinfo
			{journal} {Nature}\ }\textbf {\bibinfo {volume} {568}},\ \bibinfo {pages}
		{70} (\bibinfo {year} {2019})}\BibitemShut {NoStop}%
	\bibitem [{\citenamefont {Shen}\ \emph {et~al.}(2021)\citenamefont {Shen},
		\citenamefont {Su}, \citenamefont {Lin}, \citenamefont {Chou}, \citenamefont
		{Cheng}, \citenamefont {Park}, \citenamefont {Chiu}, \citenamefont {Lu},
		\citenamefont {Tang}, \citenamefont {Tavakoli}, \citenamefont {Pitner},
		\citenamefont {Ji}, \citenamefont {Cai}, \citenamefont {Mao}, \citenamefont
		{Wang}, \citenamefont {Tung}, \citenamefont {Li}, \citenamefont {Bokor},
		\citenamefont {Zettl}, \citenamefont {Wu}, \citenamefont {Palacios},
		\citenamefont {Li},\ and\ \citenamefont {Kong}}]{shen_ultralow_2021}%
	\BibitemOpen
	\bibfield  {author} {\bibinfo {author} {\bibfnamefont {P.-C.}\ \bibnamefont
			{Shen}}, \bibinfo {author} {\bibfnamefont {C.}~\bibnamefont {Su}}, \bibinfo
		{author} {\bibfnamefont {Y.}~\bibnamefont {Lin}}, \bibinfo {author}
		{\bibfnamefont {A.-S.}\ \bibnamefont {Chou}}, \bibinfo {author}
		{\bibfnamefont {C.-C.}\ \bibnamefont {Cheng}}, \bibinfo {author}
		{\bibfnamefont {J.-H.}\ \bibnamefont {Park}}, \bibinfo {author}
		{\bibfnamefont {M.-H.}\ \bibnamefont {Chiu}}, \bibinfo {author}
		{\bibfnamefont {A.-Y.}\ \bibnamefont {Lu}}, \bibinfo {author} {\bibfnamefont
			{H.-L.}\ \bibnamefont {Tang}}, \bibinfo {author} {\bibfnamefont {M.~M.}\
			\bibnamefont {Tavakoli}}, \bibinfo {author} {\bibfnamefont {G.}~\bibnamefont
			{Pitner}}, \bibinfo {author} {\bibfnamefont {X.}~\bibnamefont {Ji}}, \bibinfo
		{author} {\bibfnamefont {Z.}~\bibnamefont {Cai}}, \bibinfo {author}
		{\bibfnamefont {N.}~\bibnamefont {Mao}}, \bibinfo {author} {\bibfnamefont
			{J.}~\bibnamefont {Wang}}, \bibinfo {author} {\bibfnamefont {V.}~\bibnamefont
			{Tung}}, \bibinfo {author} {\bibfnamefont {J.}~\bibnamefont {Li}}, \bibinfo
		{author} {\bibfnamefont {J.}~\bibnamefont {Bokor}}, \bibinfo {author}
		{\bibfnamefont {A.}~\bibnamefont {Zettl}}, \bibinfo {author} {\bibfnamefont
			{C.-I.}\ \bibnamefont {Wu}}, \bibinfo {author} {\bibfnamefont
			{T.}~\bibnamefont {Palacios}}, \bibinfo {author} {\bibfnamefont {L.-J.}\
			\bibnamefont {Li}},\ and\ \bibinfo {author} {\bibfnamefont {J.}~\bibnamefont
			{Kong}},\ }\href {https://doi.org/10.1038/s41586-021-03472-9} {\bibfield
		{journal} {\bibinfo  {journal} {Nature}\ }\textbf {\bibinfo {volume} {593}},\
		\bibinfo {pages} {211} (\bibinfo {year} {2021})}\BibitemShut {NoStop}%
\end{thebibliography}

%

\end{document}



\title{Isolating the Nonlinear Optical Response of a \ce{MoS2} Monolayer under Extreme Screening of a Metal Substrate}
\textbf{\textit{Supplemental Material for}}

\author{Tao Yang}
\email{tao.yang@uni-due.de}
 \affiliation{Faculty of Physics, University of Duisburg-Essen, 47057 Duisburg, Germany}
\author{Stephan Sleziona}
 \affiliation{Faculty of Physics, University of Duisburg-Essen, 47057 Duisburg, Germany}
\author{Erik Pollmann}
 \affiliation{Faculty of Physics, University of Duisburg-Essen, 47057 Duisburg, Germany}
\author{Eckart Hasselbrink}
 \affiliation{Faculty of Chemistry, University of Duisburg-Essen, 45117 Essen, Germany}
\author{Peter Kratzer}
 \affiliation{Faculty of Physics, University of Duisburg-Essen, 47057 Duisburg, Germany}
\author{Marika Schleberger}
 \affiliation{Faculty of Physics, University of Duisburg-Essen, 47057 Duisburg, Germany}
\author{R. Kramer Campen}
 \affiliation{Faculty of Physics, University of Duisburg-Essen, 47057 Duisburg, Germany}
\author{Yujin Tong}
\email{yujin.tong@uni-due.de}
 \affiliation{Faculty of Physics, University of Duisburg-Essen, 47057 Duisburg, Germany}

\date{\today}

\maketitle

\section*{\hfil Contents \hfil}

\ref{fig}. Figures 

\ref{details}. Sample preparation, laser setup, and azimuthal-dependent FS-SFG measurement

\ref{approximation}. Justification of using $|\chi^{(2)}|$ to characterize the band edge

\ref{DFT}. Density functional theory calculations and density of states

Reference


\pagebreak

\section{\label{fig}Figures}

\begin{figure}[h]
\renewcommand{\thefigure}{S\arabic{figure}}
    \centering
    \includegraphics[scale=0.4]{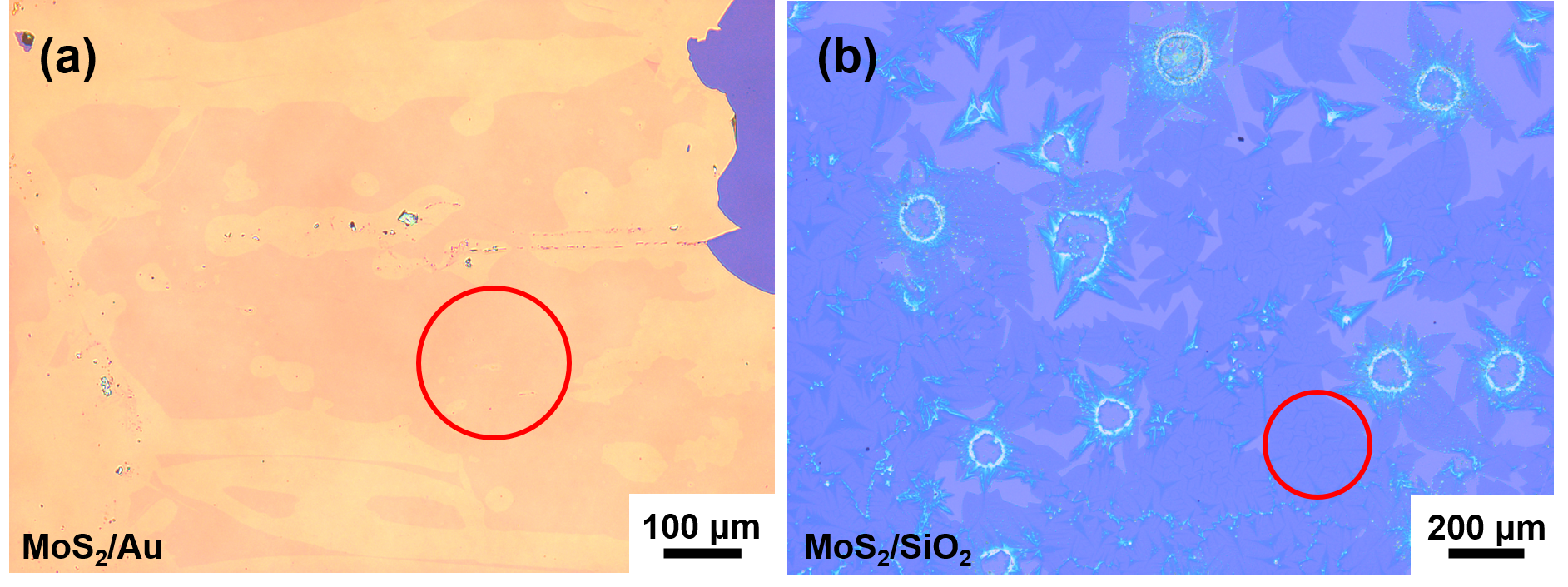}
    \caption{Optical images of (a) \ce{MoS2}/Au and (b) \ce{MoS2}/\ce{SiO2}. The region marked by the red circle is \ce{MoS2} monolayer.}
    \label{fig:S1}
\end{figure}

\begin{figure}[h]
\renewcommand{\thefigure}{S\arabic{figure}}
    \centering
    \includegraphics[scale=0.5]{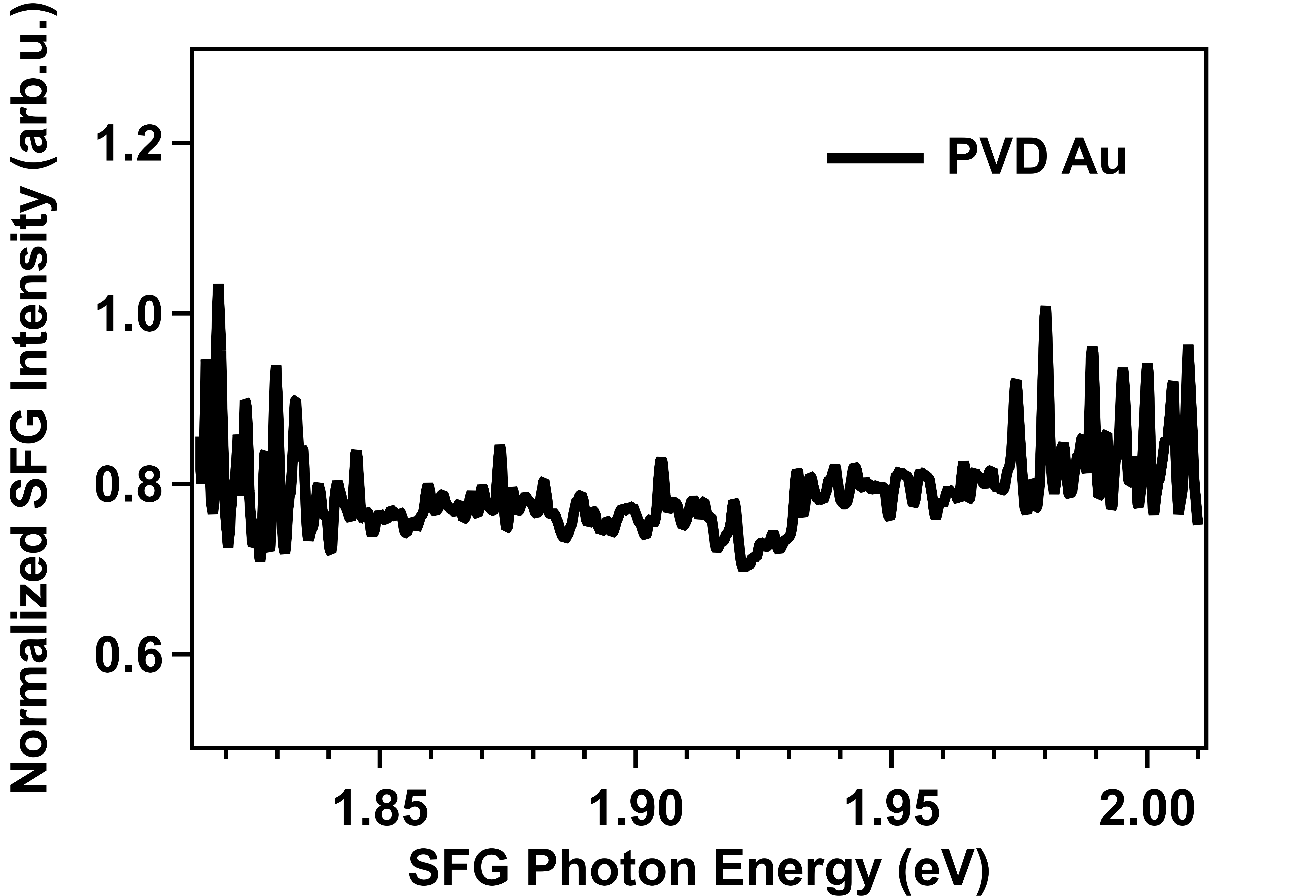}
    \caption{The normalized SFG spectrum of PVD Au collected under \textit{ppp} polarization combination.}
    \label{fig:S2}
\end{figure}

\begin{figure}[h]
\renewcommand{\thefigure}{S\arabic{figure}}
    \centering
    \includegraphics[scale=0.5]{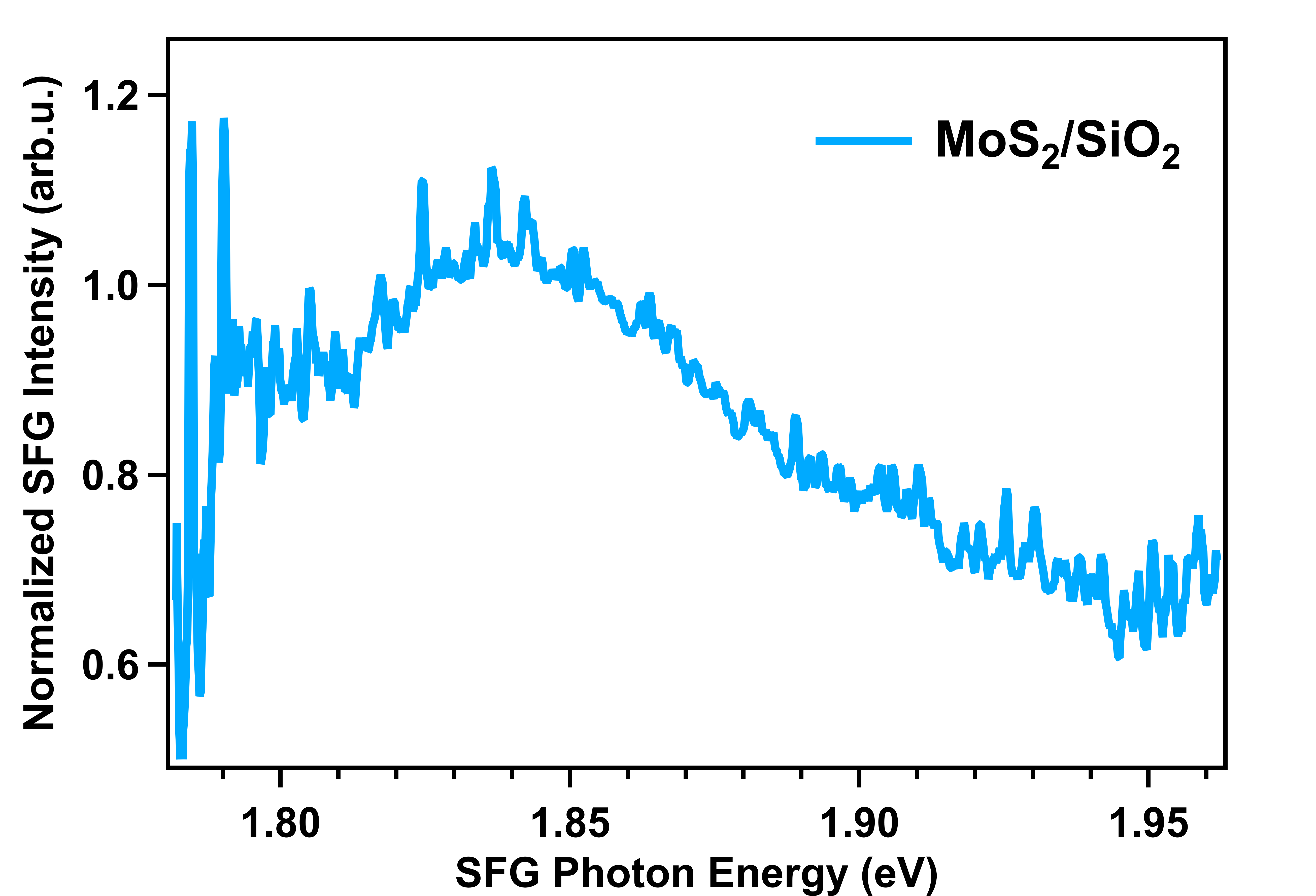}
    \caption{The normalized SFG spectrum of MoS$_2$ monolayer on SiO$_2$ collected under \textit{ppp} polarization combination.}
    \label{fig:S3}
\end{figure}

\begin{figure}[h]
\renewcommand{\thefigure}{S\arabic{figure}}
    \centering
    \includegraphics[scale=0.7]{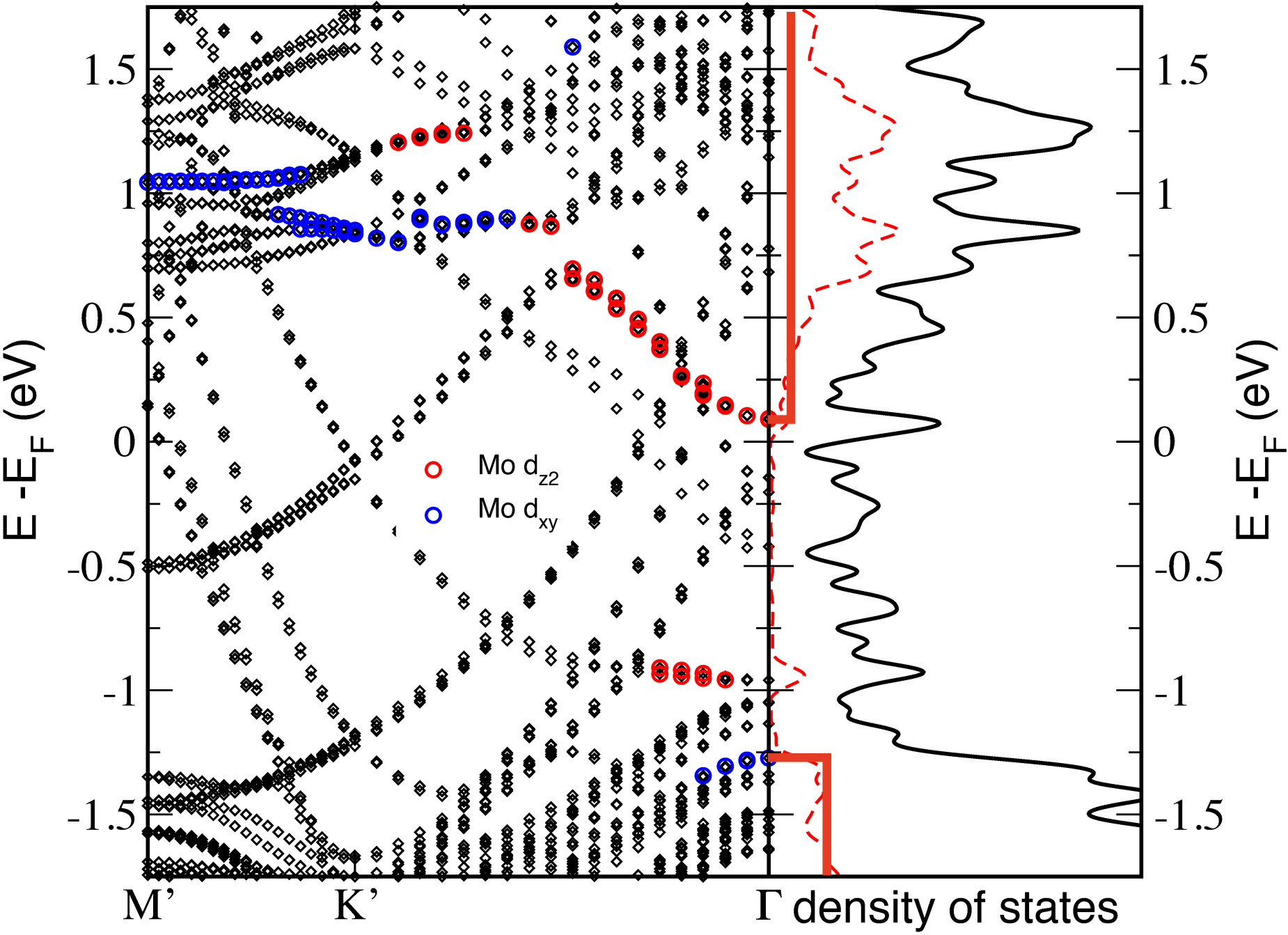}
    \caption{Density functional theory (DFT) calculated band structure and corresponding orbital projected electronic density of states (DOS) of a ($\sqrt{3}\times\sqrt{3}$) \ce{MoS2} monolayer on Au(111). The total DOS is plotted as a black solid line, the \ce{MoS2}-related DOS is plotted as a red dashed line. A step function (red solid line) can be used to describe the constant DOS of \ce{MoS2} near the band edges.}
    \label{fig:S4}
\end{figure}

\section{\label{details}Sample preparation, laser setup, and azimuthal-dependent FS-SFG measurement}
\subsection{Sample preparation}
The sample preparation is based on the commonly used micromechanical exfoliation and has been reported elsewhere\cite{pollmann_large-area_2021}. The \ce{SiO2}/Si substrate was first cleaned and stored in ethanol before use. A Ti adhesive layer (5 nm) was grown by e-beam sputtering in a PVD system followed by deposition of a Au layer (25 nm) on the clean \ce{SiO2}/Si substrate. The base pressure of the PVD-chamber was set to $1.5\times10^{-5}$ mbar during deposition. After removing the freshly produced Au surface from the vacuum chamber, we immediately performed the conventional micromechanical exfoliation of \ce{MoS2}, resulting in a large area covered with a \ce{MoS2} monolayer. 
For the preparation details of CVD-grown \ce{MoS2} monolayer on \ce{SiO2}/Si, refer to our previous study\cite{pollmann_apparent_2020}. Briefly, we employed a three-zone split tube furnace for the growth process. 
A ceramic boat with 50 mg sulfur powder is placed in the  heating zone upstream, while a separate ceramic boat containing less than 1 mg of \ce{MoO3} powder and a pre-treated \ce{SiO2} substrate is placed in the adjacent downstream heating zone. 
The furnace tube is initially purged with argon for 30 minutes, and this is maintained throughout process. 
The heating zone containing the \ce{MoO3} powder and \ce{SiO2} substrate is heated to a temperature between 750-800${^\circ}$C at a rate of $1600 {^\circ}$C/h. 
After 30 minutes, the sulfur heating zone was heated to $180 {^\circ}$C within 5 minutes. 
The furnace was held at maximum temperature for 25 minutes before being opened for rapid cooling.

\subsection{Laser setup and azimuthal-dependent FS-SFG measurement}
We utilized a laser system consisting of a Ti: Sapphire oscillator (Vitara, Coherent) and a regenerative amplifier (Legend Elite Due HE + USP, Coherent). 
A portion of the regenerative amplifier output drove a commercial optical parametric amplifier (TOPAS-Prime, Light Conversion), where the signal and idler output were combined using a collinear difference frequency generation scheme, resulting in the production of infrared (IR). 
The narrow-band visible pulse was generated using an air-spaced etalon from SLS Optics Ltd. 
Both the visible beam (0.8 $\mu$J/pulse, full width at half maximum, FWHM: 0.003 eV, 1 kHz) and the tunable infrared beam ( below 1 $\mu$J/pulse, FWHM: 0.07 eV, 1 kHz) were propagated in the \textit{X-Z} plane and had incident angles of 64$^{\circ}$ and 46$^{\circ}$, respectively. 
The sample was positioned randomly on a rotational stage, and while measuring the azimuthal dependence, the beam configuration remained the same while the sample was rotated relative to the surface normal.
The SFG signal was dispersed in a HRS-300 spectrograph from Princeton Instruments and projected onto a PI-MAX4 emICCD camera, also manufactured by Princeton Instruments.
The integrated SFG intensity was plotted as a function of azimuthal angle to guide the acquisition of FS-SFG spectrum at specific azimuthal angles, e.g., the angle with maximum intensity.
To capture a broad FS-SFG spectrum covering the energy range of A and B excitons, the center frequency of the infrared beam was set from 2900 to 4100 nm.
The background was collected using the same settings by blocking the infrared beam.
Finally, the background-free FS-SFG spectrum was added up and then normalized to the SFG spectra of our reference sample, i.e., the $\alpha$- quartz.   

\section{\label{approximation}Justification of using $|\chi^{(2)}|$ to characterize the band edge}

While the method we used here is similar to the conventional \textit{Tauc plot} based on linear optical absorption, there is also a notable difference. Here we approximate $|\chi^{(2)}|$ instead of using the imaginary part of $\chi^{(2)}$. 
Isolating the imaginary component of such an extremely broad feature necessitates heterodyne measurement, which is currently unavailable in our laboratory. 
Nevertheless, previous research has shown that when the sample system consists of both nonresonant and resonant contributions, self-heterodyne occurs due to interference between the two contributions \cite{ward_orientation_1993}. This is exactly the case for the \textit{ssp} measurement (Fig. 3 in the main text). All spectra clearly show a linear response, regardless of the relative phases between the Au contribution (which can be considered a nonresonance) and that of the \ce{MoS2} monolayer. These results clearly confirm that the lineshape of the imaginary part of $\chi^{(2)}$ of \ce{MoS2} on Au substrate is also linear, justifying the validity of the analogous \textit{Tauc plot} (Fig. 4 in the main text).

\section{\label{DFT}Density functional theory calculations and density of states}

Electronic structure calculations of a \ce{MoS2} monolayer adsorbed on a Au(111) surface were carried out using density functional theory (DFT).
The generalized-gradient approximation GGA-PBE was used to describe electronic exchange and correlation effects.
The computational settings were similar to previous work \cite{sarkar_signatures_2021}, but spin-orbit coupling has been included by performing a DFT calculation allowing for non-collinear spin structure.
The band structure (left panel) and corresponding orbital projected electronic density of states (DOS) (right panel) of a ($\sqrt{3}\times\sqrt{3}$) \ce{MoS2} monolayer on Au(111) are shown in Fig.\ref{fig:S4}. 
The labelling of the high-symmetry points refers to the ($\sqrt{3}\times\sqrt{3}$) unit cell, i.e., the K point of the commonly discussed ($1\times1$) unit cell got backfolded onto the $\Gamma$ point of the supercell. 
Therefore, we discuss optical transitions at the $\Gamma$ point.
The total band structure is traced out by the black rhombuses, whereas the red and blue circles mark those bands that have a significant contribution from $4d_{z^2}$ and $4d_{xy}$ orbitals of the Mo atoms, respectively. 
It should be note that the optical de-excitation takes place between the conduction band states that result from hybridization of the Mo $4d_{z^2}$ orbitals with sulphur orbitals and the valence band states of Mo $4d_{xy}$ character.
Therefore the DFT calculation suggests a Kohn-Sham band gap (accessible to optical transitions) of $\sim1.3$ eV for a \ce{MoS2} monolayer on Au(111).
In the right panel of Fig.\ref{fig:S4}, the black solid line and red dashed line illustrate the total DOS and projected DOS onto the Mo 4d orbitals, respectively.
The red solid line depicts by step functions the constant DOS of \ce{MoS2} near the edges of the involved valence and conduction bands.
At energies further away from the band edges, other bands start to contribute and the DOS is no longer a constant.


%